\documentclass[aps,prl,twocolumn,showpacs,superscriptaddress,nobibnotes,10pt]{revtex4-1}
\usepackage{amsmath,amssymb,graphicx}
\usepackage{dcolumn}
\usepackage{braket}
\usepackage{latexsym}
\usepackage{mathrsfs}
\usepackage{arydshln}
\usepackage{epsfig}
\usepackage{epstopdf}
\usepackage{natbib}
\usepackage{color}
\newcommand{\bitem}{\begin{itemize}}
\newcommand{\fitem}{\end{itemize}}
\newcommand{\beq}{\begin{equation}}
\newcommand{\eeq}{\end{equation}}
\newcommand{\beqa}{\begin{eqnarray}}
\newcommand{\eeqa}{\end{eqnarray}}

\begin{document}

\title{Cooperative spontaneous emission via renormalization approach: Classical versus semi-classical effects}

\author{C. E. M\'aximo} \affiliation{Departamento de F\'{i}sica, Universidade Federal de S\~ao Carlos, P.O. Box 676, 13565-905, São Carlos, São Paulo, Brazil}

\author{R. Bachelard} \affiliation{Departamento de F\'{i}sica, Universidade Federal de S\~ao Carlos, P.O. Box 676, 13565-905, São Carlos, São Paulo, Brazil}
\affiliation{Universit\'{e} C\^{o}te d'Azur, CNRS, INPHYNI, France}

\author{F. E. A. dos Santos} \affiliation{Departamento de F\'{i}sica, Universidade Federal de S\~ao Carlos, P.O. Box 676, 13565-905, São Carlos, São Paulo, Brazil}

\author{C. J. Villas-Boas} \affiliation{Departamento de F\'{i}sica, Universidade Federal de S\~ao Carlos, P.O. Box 676, 13565-905, São Carlos, São Paulo, Brazil}

\date{\today}
\begin{abstract}
We address the many-atom emission of a dilute cloud of two-level atoms through a renormalized perturbation theory. An analytical solution for the truncated coupled-dipole equations is derived, which contains an effective spectrum associated to the initial conditions. Our solution is able to distinguish precisely classical from semi-classical predictions for large atomic ensembles. This manifests as a reduction of the cooperativity in the radiated power for higher atomic excitation, in disagreement with the fully classical prediction from linear optics. Moreover, the second-order cooperative emission appears accurate over several single atom lifetimes and for interacting regimes stronger than those permitted in conventional perturbation theory.  We can compute the semi-classical dynamics of hundred thousand of interacting atoms with ordinary computational resources, which makes our formalism particularly promising to probe the nonlinear dynamics of quantum many-body systems that emerge from cumulant expansions.
\end{abstract}
\maketitle

Non-equilibrium collective phenomena can manifest in classical and quantum many-body systems. A particularly challenging task is the segregation of those with purely non-classical nature. Indeed, the description of many-body systems through a fully quantum model suffers from the exponential growth of the number of quantum correlations with the number of particles.  Hence, exact numerical methods, such as diagonalization, are limited to a few tens of particles, even for supercomputers. This is the reason why truncation methods were developed, based on cumulant expansions~\cite{helmut2015,1rey2015, 2rey2015,kastner2016, bachelard2017,rey2017}, in which higher-order connected correlations are neglected to obtain a more favorable scaling of the complexity with the number of particles. As a consequence, nonlinearities emerge from the decomposition of higher-order correlations into products of lower-order ones, which turns even trickier predictions on collective quantum phenomena.

At the lowest order of the cumulant expansion one can find the ``semi-classical model'', sometimes called ``mean-field approximation''. It can be derived by neglecting two-body and higher-order connected correlations, keeping the contribution from ``single-body'' auto-correlations. As a result, a set of nonlinear coupled equations is obtained only for one-operator expected values, that scales linearly with the number of particles~\cite{rey2016}. Even in this simpler case, numerical solutions for real physical systems demand long running times, due to the nonparallelizable nature of the numerical algorithms. Thus, approximated analytical solutions which allow for independent calculations might be the unique possible alternative.

In this letter, we propose to treat analytically the many-atom spontaneous emission, in the semi-classical approximation, using a renormalization method in perturbation theory. This method was explored in the nineteenth century when it was discovered that naive perturbation theory was giving rise to unphysical behavior in celestial mechanics problems~\cite{lindstedt1883}. A proposed solution to this plight arose from the so-called Poincaré-Lindstedt method~\cite{book1}, also known as renormalization method. It consists in eliminating unphysical secular terms from perturbative solutions, by renormalizing the characteristic eigenvalues of the differential equations. As a consequence, great advantages can be obtained over conventional power series solutions.  More recently, this approach has been used with variational methods~\cite{schanz2003} and applied to dynamics of Bose-Einstein condensates~\cite{pelster2011,pelster2013}.

In general, cooperative spontaneous emission presents all features of a truly quantum many-body problem. Yet the accelerated emission of radiation, known as superradiance, was understood to be essentially a semi-classical cooperative effect, as the radiative decay of a nearly fully inverted system is well described by the Bloch-Maxwell equations \cite{arecchi1972,macgillivray1976}. Only almost four decades later, a similar phenomenon was revealed in the low excitation regime~\cite{wodkiewicz2006,kurizki2007}, described by a fully-classical Coupled-Dipole Model (CDM)~\cite{1scully2008,1kaiser2010,scully2010,kaiser2011,kaiser2012,scully2015,cottier2018}, with experimental verifications provided in the linear-optics regime~\cite{1kaiser2010,felinto2014,guerin2016,roof2016}. Long-lived radiation modes, know as subradiance, were equally well captured by the CDM~\cite{eberly2006,1kaiser2012,kaiser2016,cottier2018}. Thus, the divergence between classical and semi-classical cooperative effects represents an opportunity to demonstrate the efficiency of the renormalization method.   

A natural consequence of eigenvalues renormalization is the rising of an analytical effective eigenspectrum, with strong dependence on initial conditions. We demonstrate that the average many-atom decay rate of this spectrum scales with an excitation-dependent cooperativity parameter, which is a modification of the linear-optics optical thickness. As a result, super- and subradiance are substantially reduced when the excited state population of the atoms is increased. While such cooperativity suppression cannot be predicted by the classical model, it is accurately captured by a first-order renormalized solution for the semi-classical equations.  

Let us first introduce classical and semi-classical models which describe the spontaneous emission from $N$ two-level atoms . This emission process can be monitored by the total power radiated throughout a sphere of radius $r$,
\begin{equation}
P\left(t\right)\propto\int_{0}^{2\pi}\int_{0}^{\pi}\Bigl\langle\hat{E}^{\dagger}\left(\mathbf{r},t\right)\hat{E}\left(\mathbf{r},t\right)\Bigr\rangle r^{2}d\Omega,\label{P}
\end{equation}
with $d\Omega$ an infinitesimal solid angle. In the far-field limit, the scalar electric field operator $\hat{E}\left(\hat{\mathbf{n}},t\right)$,  at the observation point $\hat{\mathbf{n}}$, relates to the radiating dipoles as
\begin{equation}
\hat{E}\left(\hat{\mathbf{n}},t\right)\propto\frac{1}{r}\sum_{m=1}^{N}e^{-ik_{0}\hat{\mathbf{n}}\cdot\mathbf{r}_{m}}\hat{\sigma}_{m}^{-}\left(t\right),\label{E}
\end{equation}
where  $\mathbf{r}_m$ are the constant center-of-mass position of the atoms, randomly distributed in the three-dimensional space. $\hat{\sigma}_{m}^{-}=\bigl|g_{m}\bigr\rangle\bigl\langle e_{m}\bigr|$ and $\hat{\sigma}_{m}^{+}=\left(\hat{\sigma}_{m}^{-}\right)^{\dagger}$
correspond, respectively, to the lowering and rising dipole operators, with $\bigl|g_{m}\bigr\rangle$ ( $\bigl|e_{m}\bigr\rangle$) the ground (excited) state. For the derivation of Eq.\eqref{E}, one considers the standard atom-light interaction, with a continuous integration over all vacuum modes of the electromagnetic field~\cite{kaiser2011,kaiser2012}, together with the Markov and rotating wave approximations. We assume that the atomic cloud is previously prepared by a laser field, with wave vector $\mathbf{k}_{0}=k_{0}\hat{\mathbf{z}}$ and wavelength $\lambda=2\pi/k_0$, detuned by $\Delta$ from the atomic resonance. We then observe the atomic dipoles spontaneously decaying after switching off the pump.

By substituting Eq.\eqref{E} in Eq.\eqref{P}, the radiated power can be expressed as a sum over two-atom quantum correlators.  Therefore, a solution for the set of $2^{N}$ coupled differential equations, that contains $N$-body correlations between atoms, is required. In order to prevent such exponential growth, one typically truncates the system of equations up to some order in connected correlations~\cite{Kramer2015, Young2018}. In this work, we neglect the correlations between the quantum fluctuations of different atomic dipoles ($\bigl\langle\hat{\sigma}_{m}^{\alpha}\hat{\sigma}_{n}^{\beta}\bigr\rangle\approx\bigl\langle\hat{\sigma}_{m}^{\alpha}\bigr\rangle\bigl\langle\hat{\sigma}_{n}^{\beta}\bigr\rangle$, for $m\neq n$), keeping the single-body correlations of equal indexes ($\bigl\langle\hat{\sigma}_{m}^{+}\hat{\sigma}_{m}^{-}\bigr\rangle=\left(1+\bigl\langle\hat{\sigma}_{m}^{z}\bigr\rangle\right)/2$),  with $\hat{\sigma}_{m}^{z}=\bigl|e_{m}\bigr\rangle\bigl\langle e_{m}\bigr|-\bigl|g_{m}\bigr\rangle\bigl\langle g_{m}\bigr|$ the inversion population operator. This is the origin of the name ``semi-classical''. 

Using this approximation, the total power \eqref{P} reduces to
\begin{equation}
P_{\mathrm{sc}}\propto\sum_{m=1}^{N}\biggl[\frac{1+\bigl\langle\hat{\sigma}_{m}^{z}\bigr\rangle}{2}+\sum_{n\neq m}\Re\left(u_{mn}\bigl\langle\hat{\sigma}_{m}^{+}\bigr\rangle\bigl\langle\hat{\sigma}_{n}^{-}\bigr\rangle\right)\biggr], \label{P_nlin}
\end{equation}
where the single-atom contributions have been explicitly separated from the second sum. The time evolution of the total power is then completely determined by solving the following set of nonlinear equations:
\begin{subequations} 
\begin{eqnarray}
\overset{.}{\bigl\langle\hat{\sigma}_{m}^{-}\bigr\rangle} &=& \left(i\Delta-\frac{\Gamma}{2}\right)\bigl\langle\hat{\sigma}_{m}^{-}\bigr\rangle+\frac{\Gamma}{2}\sum_{n\neq m}u_{mn}\bigl\langle\hat{\sigma}_{m}^{z}\bigr\rangle\bigl\langle\hat{\sigma}_{n}^{-}\bigr\rangle ,\label{sig1_av} \\
\overset{.}{\bigl\langle\hat{\sigma}_{m}^{z}\bigr\rangle} &=& -\Gamma\left(1+\bigl\langle\hat{\sigma}_{m}^{z}\bigr\rangle\right)-2\Gamma\sum_{n\neq m}\Re\left(u_{mn}\bigl\langle\hat{\sigma}_{m}^{+}\bigr\rangle\bigl\langle\hat{\sigma}_{n}^{-}\bigr\rangle\right), \label{sigz_av}
\end{eqnarray}\label{sig_nlin}
\end{subequations}
whose number of variables increases linearly with $N$. For the derivation of Eqs.(\ref{P_nlin}-\ref{sig_nlin}),  the same conditions used to derive Eq.\eqref{E} are assumed~\cite{rey2016,bachelard2017}. In particular, the single atom decay rate $\Gamma$ and the light-mediated long-range interactions $u_{mn}\equiv\exp\left(ik_{0}\bigl|\mathbf{r}{}_{m}-\mathbf{r}{}_{n}\bigr|\right)/ik_{0}\bigl|\mathbf{r}{}_{m}-\mathbf{r}{}_{n}\bigr|$ arise, by first principles, from the continuous integration over all vacuum modes of the electromagnetic field. The effective interactions $u_{mn}$ are valid in the optically dilute regime, where near-field terms do not affect the scattering. 

In the low excitation regime, all the atoms evolve close to the ground state ($\bigl\langle\hat{\sigma}_{m}^{z}\bigr\rangle\approx-1$), and even the single-atom correlations separate: $\bigl\langle\hat{\sigma}_{m}^{+}\hat{\sigma}_{m}^{-}\bigr\rangle\approx\bigl\langle\hat{\sigma}_{m}^{+}\bigr\rangle\bigl\langle\hat{\sigma}_{m}^{-}\bigr\rangle$. In such case, Eqs.(\ref{P_nlin}-\ref{sig_nlin}) reduce even further, giving rise to the linear-optics CDM, that can be derived from the classical Maxwell equations~\cite{scully2010,cottier2018}.

Regarding the initial conditions, we hereafter assume a mean-field-like initial state,
\begin{equation}
\left(\bigl\langle\hat{\sigma}_{m}^{-}\left(0\right)\bigr\rangle,\bigl\langle\hat{\sigma}_{m}^{z}\left(0\right)\bigr\rangle\right)=\left(e^{-i\mathbf{k}_{0}\cdot\mathbf{r}_{m}}\beta,2p_{e}-1\right),\label{ini}
\end{equation}
where all atoms have an equal probability $p_e$ of being excited by the laser. However, different phases $\mathbf{k}_{0}\cdot\mathbf{r}_{m}$ are kept for the coherences of the atoms, which are proportional to the real amplitude $\beta=\sqrt{p_e\left(1-p_e\right)}$. In the low excitation regime ($p_e \rightarrow 0$), the classical case is recovered ($\beta \approx \sqrt{p_e}$). Oppositely, in the strong excitation limit ($p_e \rightarrow 1$), the coherence vanishes ($\beta \rightarrow 0$).

In order to extract an approximate solution from Eq.\eqref{sig_nlin}, let us consider a broader class of inhomogeneous nonlinear differential equations,
\begin{equation}
\dot{x}_{l}=-V_{l}-M_{lm}x_{m}-T_{lmn}x_{m}x_{n}, \label{dx}
\end{equation}
with $\textbf{V}$ a complex vector, $\textbf{M}$ a non-Hermitian diagonalizable matrix and $\textbf{T}$ a  complex tensor, all of them constant in time. Hereafter, the Einstein summation rule is adopted for Latin indices but not for the Greek ones. We then transform Eq.\eqref{dx} into a integral equation:
\begin{multline}
x_{l}(t)=\int_{0}^{\infty}dt^{\prime}G_{lm}(t,t^{\prime})V_{m}-G_{lm}(t,0)x_{m}(0)\\
+\int_{0}^{\infty}dt^{\prime}G_{lm}\left(t,t^{\prime}\right)T_{mno}x_{n}\left(t^{\prime}\right)x_{o}\left(t^{\prime}\right),\label{x}
\end{multline}
where the propagator $G_{lm}(t,t^{\prime})$ satisfies Green's linear equation
\begin{equation}
\frac{dG_{lm}}{dt}+M_{ln}G{}_{nm}=-\delta\left(t-t^{\prime}\right)\delta_{lm}.\label{dG}
\end{equation}

In the Supplementary Material (SM), we prove that the solution,
\begin{equation}
G_{lm}\left(t-t^{\prime}\right)=-\sum_{\alpha}\psi_{l}^{\left(\alpha\right)}\phi_{m}^{\left(\alpha\right)}\theta\left(t-t^{\prime}\right)e^{-\gamma^{\left(\alpha\right)}\left(t-t^{\prime}\right)}, \label{G}
\end{equation}
satisfies Eq.\eqref{dG}, where $\gamma^{\left(\alpha\right)}$ and $\psi_{m}^{\left(\alpha\right)}$ are, respectively, the eigenvalues and eigenvectors of $M_{lm}$, with $\theta\left(t-t^{\prime}\right)$  the Heaviside step function. We have also defined the vector
\begin{equation}
\phi_{l}^{\left(\alpha\right)} \equiv \sum_{\beta}\left[\mathbf{J}^{-1}\right]^{\left(\alpha\beta\right)}\psi_{l}^{\left(\beta\right)},\label{phi}
\end{equation}
with $\left[\mathbf{J}\right]^{\left(\alpha\beta\right)} \equiv \psi_{l}^{\left(\alpha\right)}\psi_{l}^{\left(\beta\right)}$. $\psi_{l}^{\left(\alpha\right)}$ and $\phi_{m}^{\left(\alpha\right)}$ satisfy the closure relation $\sum_{\alpha}\psi_{l}^{\left(\alpha\right)}\phi_{m}^{\left(\alpha\right)}=\delta_{lm}$,  and no orthogonality condition is assumed between them (see SM).

Substituting \eqref{G} in the integral equation (\ref{x}), we obtain the more explicit expression:
\begin{multline}
x_{l}(t)=x_{l}^{(0)}(t)-\sum_{\alpha}\psi_{l}^{\left(\alpha\right)}\phi_{m}^{\left(\alpha\right)}T_{mno}e^{-\gamma^{\left(\alpha\right)}t}\\
\times\int_{0}^{t}dt^{\prime}x_{n}\left(t^{\prime}\right)x_{o}\left(t^{\prime}\right)e^{\gamma^{\left(\alpha\right)}t^{\prime}},\label{xx}
\end{multline}
where we have found analytically the zeroth-order solution
\begin{equation}
x_{l}^{(0)}(t)\equiv \sum_{\alpha}\psi_{l}^{\left(\alpha\right)}\left(b^{\left(\alpha\right)}e^{-\gamma^{\left(\alpha\right)}t}-a^{\left(\alpha\right)}\right), \label{x0}
\end{equation}
with $a^{\left(\alpha\right)}\equiv \phi_{m}^{\left(\alpha\right)}V_{m}/\lambda^{\left(\alpha\right)}$ the steady-state amplitude and $b^{\left(\alpha\right)}\left(0\right)\equiv a^{\left(\alpha\right)} + \phi_{m}^{\left(\alpha\right)}x_{m}\left(0\right)$ the transient amplitude.

The next step is to iterate solution \eqref{xx}, keeping up to first-order terms in $T_{mno}$. As a result, the variables $x_{n}\left(t^{\prime}\right)$ and $x_{o}\left(t^{\prime}\right)$, in Eq.\eqref{xx}, are replaced by the zeroth-order expression \eqref{x0}. The total solution can approximately be written as $x_{l}(t)\approx x_{l}^{(0)}(t)+x_{l}^{(1)}(t)$, where we define the first-order temporal solution
\begin{multline}
x_{l}^{\left(1\right)}\left(t\right)=-\sum_{\alpha}\psi_{l}^{\left(\alpha\right)}e^{-\gamma^{\left(\alpha\right)}t}\Biggl[A^{\left(\alpha\right)}\int_{0}^{t}dt^{\prime}e^{\gamma^{\left(\alpha\right)}t^{\prime}}\\
-\sum_{\beta}B^{\left(\alpha \beta \right)}\int_{0}^{t}dt^{\prime}e^{\left(\gamma^{\left(\alpha\right)}-\gamma^{\left(\beta\right)}\right)t^{\prime}}\\
+\sum_{\beta,\delta}C^{\left(\alpha \beta \delta \right)}\int_{0}^{t}dt^{\prime}e^{\left(\gamma^{\left(\alpha\right)}-\gamma^{\left(\beta\right)}-\gamma^{\left(\delta\right)}\right)t^{\prime}}\Biggr]. \label{x1}
\end{multline}
The terms proportional to the tensor components $T_{mno}$ have been absorbed into the following coefficients:
\begin{subequations} 
\begin{eqnarray}
A^{\left(\alpha\right)}&\equiv &\sum_{\beta \delta}\phi_{m}^{\left(\alpha\right)}T_{mno}\psi_{n}^{\left(\beta\right)}a^{\left(\beta\right)}\psi_{o}^{\left(\delta\right)}a^{\left(\delta\right)},\label{A}\\
B^{\left(\alpha \beta\right)}&\equiv &\sum_{\delta}\phi_{m}^{\left(\alpha\right)}\left(T_{mno}\psi_{n}^{\left(\delta\right)}a^{\left(\delta\right)}\psi_{o}^{\left(\beta\right)}b^{\left(\beta\right)}\right. \nonumber \\&&\qquad\quad\left.+T_{mon}\psi_{o}^{\left(\beta\right)}b^{\left(\beta\right)}\psi_{n}^{\left(\delta\right)}a^{\left(\delta\right)}\right),\label{B}\\
C^{\left(\alpha \beta \delta\right)}&\equiv &\phi_{m}^{\left(\alpha\right)}T_{mno}\psi_{n}^{\left(\beta\right)}b^{\left(\beta\right)}\psi_{o}^{\left(\delta\right)}b^{\left(\delta\right)}.\label{C}
\end{eqnarray}\label{coff}
\end{subequations}
Note that no assumption has been made regarding the nature of the generalized interactions $T_{mno}$.

So far we have performed nothing more than what one could call naive perturbation theory. The main problem of simply applying the iterative procedure is the appearance of unphysical secular terms in the perturbative solution. This becomes clear when the time integrals in Eq.\eqref{x1} are analytically solved, after separating the cases $\gamma^{\left(\alpha\right)}=\gamma^{\left(\beta\right)}$ and $\gamma^{\left(\alpha\right)}=\gamma^{\left(\beta\right)}+\gamma^{\left(\delta\right)}$ from $\gamma^{\left(\alpha\right)}\neq\gamma^{\left(\beta\right)}$ and $\gamma^{\left(\alpha\right)}\neq\gamma^{\left(\beta\right)}+\gamma^{\left(\delta\right)}$. The result at first order reads:
\begin{multline}
x_{l}\left(t\right)\approx \sum_{\alpha}\psi_{l}^{\left(\alpha\right)}\Biggl[\left(b^{\left(\alpha\right)}+tS^{\left(\alpha\right)}\right)e^{-\gamma^{\left(\alpha\right)}t}-a^{\left(\alpha\right)}\\
-A^{\left(\alpha\right)}\frac{1-e^{-\gamma^{\left(\alpha\right)}t}}{\gamma^{\left(\alpha\right)}}-\overset{{\scriptscriptstyle \left(\gamma^{\left(\alpha\right)}\neq\gamma^{\left(\beta\right)}\right)}}{\sum_{\beta}}B^{\left(\alpha\beta\right)}\frac{e^{-\gamma^{\left(\alpha\right)}t}-e^{-\gamma^{\left(\beta\right)}t}}{\gamma^{\left(\alpha\right)}-\gamma^{\left(\beta\right)}}\\
+\overset{{\scriptscriptstyle \left(\gamma^{\left(\alpha\right)}\neq\gamma^{\left(\beta\right)}+\gamma^{\left(\delta\right)}\right)}}{\sum_{\beta,\delta}\quad C^{\left(\alpha\beta\delta\right)}}\frac{e^{-\gamma^{\left(\alpha\right)}t}-e^{-\left(\gamma^{\left(\beta\right)}+\gamma^{\left(\delta\right)}\right)t}}{\gamma^{\left(\alpha\right)}-\gamma^{\left(\beta\right)}-\gamma^{\left(\delta\right)}}\Biggr],\label{xx01}
\end{multline}
where a linear growth in time appears in the first line of Eq.\eqref{xx01}, multiplying the secular coefficient
\begin{equation}
S^{\left(\alpha\right)}\equiv\overset{{\scriptscriptstyle \left(\gamma^{\left(\alpha\right)}=\gamma^{\left(\beta\right)}\right)}}{\sum_{\beta}}B^{\left(\alpha\beta\right)}-\overset{{\scriptscriptstyle \left(\gamma^{\left(\alpha\right)}=\gamma^{\left(\beta\right)}+\gamma^{\left(\delta\right)}\right)}}{\sum_{\beta,\delta}\quad C^{\left(\alpha\beta\delta\right)}}.\label{S}
\end{equation}
The temporal behavior $te^{-\gamma^{\left(\alpha\right)}t}$ is problematic for Markovian dissipative systems since it represents an energy gain for short times. Such type of growth might even turn first-order terms larger than zeroth-order ones.

In order to eliminate secular terms from the dynamics, we resort to the Poincar\'e-Lindstedt method ~\cite{book1}. It consists in rewriting the eigenvalues in the form $\gamma^{\left(\alpha\right)}=\bar{\gamma}^{\left(\alpha\right)}+\delta\gamma^{\left(\alpha\right)}$, where we define the renormalized eigenspectrum $\bar{\gamma}^{\left(\alpha\right)}\equiv\gamma^{\left(\alpha\right)}-\delta\gamma^{\left(\alpha\right)}$. The frequency $\delta\gamma^{\left(\alpha\right)}$ is a mathematical artifice to be calculated later. We now replace all exponentials in Eq.\eqref{xx01} by the series expansion
\begin{equation}
e^{-\gamma^{\left(\alpha\right)}t}=e^{-\bar{\gamma}^{\left(\alpha\right)}t}\left(1-\delta\gamma^{\left(\alpha\right)}t+\cdots\right),\label{expan}
\end{equation}
taking care of neglecting second-order terms and higher in $\delta\gamma^{\left(\alpha\right)}$. It is straightforward to realize that the choice $\delta\gamma^{\left(\alpha\right)}=S^{\left(\alpha\right)}/b^{\left(\alpha\right)}$ cancels out the linear growth, and the following expression is finally obtained:
\begin{multline}
x_{l}\left(t\right)\approx\sum_{\alpha}\psi_{l}^{\left(\alpha\right)}\Biggl[b^{\left(\alpha\right)}e^{-\bar{\gamma}^{\left(\alpha\right)}t}-a^{\left(\alpha\right)}-A^{\left(\alpha\right)}\frac{1-e^{-\bar{\gamma}^{\left(\alpha\right)}t}}{\gamma^{\left(\alpha\right)}}\\-\overset{{\scriptscriptstyle \left(\gamma^{\left(\alpha\right)}\neq\gamma^{\left(\beta\right)}\right)}}{\sum_{\beta}}B^{\left(\alpha\beta\right)}\frac{e^{-\bar{\gamma}^{\left(\alpha\right)}t}-e^{-\bar{\gamma}^{\left(\beta\right)}t}}{\gamma^{\left(\alpha\right)}-\gamma^{\left(\beta\right)}}\\+\overset{{\scriptscriptstyle \left(\gamma^{\left(\alpha\right)}\neq\gamma^{\left(\beta\right)}+\gamma^{\left(\delta\right)}\right)}}{\sum_{\beta,\delta}\quad C^{\left(\alpha\beta\delta\right)}}\frac{e^{-\bar{\gamma}^{\left(\alpha\right)}t}-e^{-\left(\bar{\gamma}^{\left(\beta\right)}+\bar{\gamma}^{\left(\delta\right)}\right)t}}{\gamma^{\left(\alpha\right)}-\gamma^{\left(\beta\right)}-\gamma^{\left(\delta\right)}}\Biggr].\label{xren}
\end{multline}

In Eq.\eqref{xren}, each rectified exponential $e^{-\bar{\gamma}^{\left(\alpha\right)}t}$ is equivalent to the sum over the contribution of an infinite number of terms, coming from all perturbative orders, representing thus an immense improvement over naive perturbative solutions. It is worth mentioning that the new analytical eigenspectrum $\bar{\gamma}^{\left(\alpha\right)}$ does depend on the initial conditions through amplitudes $b^{\left(\alpha\right)}\left(0\right)$, a characteristic absent in linear algebra approaches.

In order to clearly understand the implications of our general formalism, let us consider the semi-classical Eqs.\eqref{sig_nlin} for cooperative spontaneous emission as a particular case of Eq.\eqref{dx}. Replacing the general components $x_l$ by the expectation values $\bigl\langle\sigma_{l}^{\alpha}\bigr\rangle$, for $\alpha=\pm,z$, $\textbf{M}$ becomes a $3N\times 3N$ diagonal matrix, where the diagonal block of three single-atom eigenvalues, $\left(\Gamma/2 - \Delta i,\Gamma/2 + \Delta i,\Gamma\right)$, is the same for all atoms. Hence the associated eigenvectors correspond to the well-known canonical basis, which reduces the general expression~\eqref{xren} to
\begin{subequations}
\begin{eqnarray}
\bigl\langle\sigma_{l}^{-}\left(t\right)\bigr\rangle&\approx&\sqrt{p_{e}\left(1-p_{e}\right)}\biggl[\left(1+p_{e}\eta_{l}\right)e^{-i\mathbf{k}_{0}\cdot\mathbf{r}_{l}-\bar{\gamma}_{-}^{\left(l\right)}t} \nonumber \\
&& -p_{e}\sum_{m\neq l}u_{lm}e^{-i\mathbf{k}_{0}\cdot\mathbf{r}_{m}-\left(\bar{\gamma}_{z}^{\left(l\right)}+\bar{\gamma}_{-}^{\left(m\right)}\right)t}\biggr],\label{analy1}\\
\bigl\langle\sigma_{l}^{z}\left(t\right)\bigr\rangle & \approx & 2p_{e}e^{-\bar{\gamma}_{z}^{\left(l\right)}t}-1.\label{analy3}
\end{eqnarray}\label{analy}
\end{subequations}
where we have introduced the semi-classical eigenspectrum at first-order
\begin{subequations}
\begin{eqnarray}
\bar{\gamma}_{\pm}^{\left(l\right)}&=&\frac{\Gamma}{2}\left[1+\Re\left(\eta_{l}\right)\right]\pm i\left[\Delta-\frac{\Gamma}{2}\Im\left(\eta_{l}\right)\right],\label{gren1}\\
\bar{\gamma}_{z}^{\left(l\right)}&=&\Gamma\left[1+\left(1-p_e\right)\Re\left(\eta_{l}\right)\right],\label{gren3}
\end{eqnarray}\label{gamma_ren}
\end{subequations}
and the structure factor
\begin{equation}
\eta_{l}\equiv\sum_{m\neq l}u_{lm}e^{i\mathbf{k}_{0}\cdot(\mathbf{r}_{l}-\mathbf{r}_{m})}.\label{eta}
\end{equation}
Note that the equations above incorporate the initial conditions~ \eqref{ini}. Moreover, as a consequence of the renormalization protocol, $\eta_l$ and the excited population $p_e$ bring cooperative corrections to the independent atoms eigenvalues.

\begin{figure}[h!]	
	\begin{center}
		\includegraphics[scale=0.435]{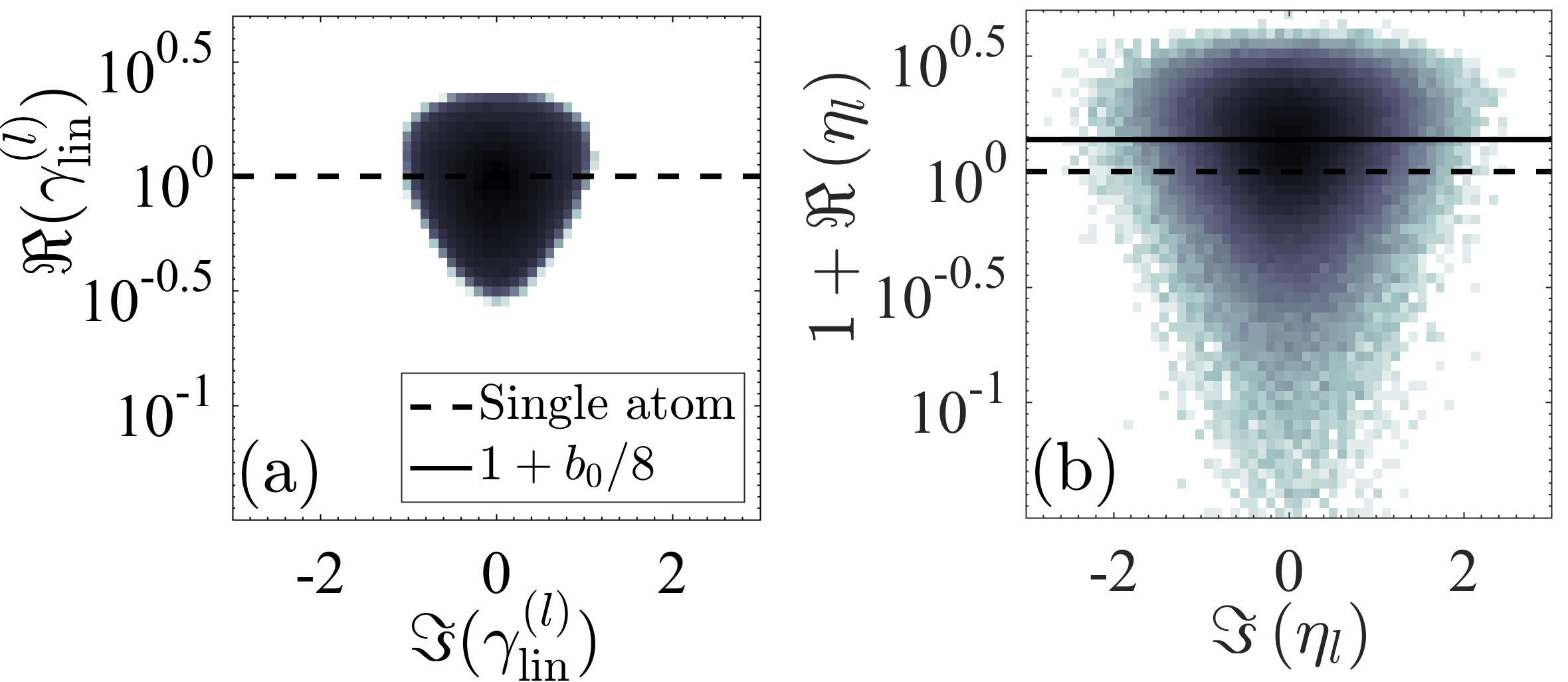}
	\end{center}
	\caption{\label{fig1} Exact dimensionless eigenspectrum $\gamma_\mathrm{lin}^{\left(l\right)}$ of the (a) classical model versus (b) the effective spectrum predicted by Eq.\eqref{gren1}, in the low-excitation regime $p_e=0.01$. Black pixels indicate many eigenvalues, whereas white pixels present zero eigenvalues. Simulations realized for $N=2000$ particles and a radius $R=5.8\lambda$ ($b_0=3$ and $\mathcal{C}=0.37$).}
\end{figure}

Let us now turn our attention to the linear portion of the analytical eigenspectrum, in Eq.\eqref{gren1}. Note that the dimensionless decay rates $1+\Re\left(\eta_{l}\right)$ and energy shifts $\Im\left(\eta_{l}\right)$ contain information on the initial state \eqref{ini}, where a laser induces phases in the dipoles (see Eq.\eqref{eta}). Thus, even in the classical regime ($p_e\ll 1$), the effective spectrum \eqref{gren1} differs substantially from the one extracted by diagonalizing the linear optics equations. We compare both in Fig.\ref{fig1} for a Gaussian atomic cloud of standard deviation $R$, with a degree of excitation $p_e=0.01$. For the exact linear-optics eigenvalues, the average of the decay rates is the single-atom linewidth for any initial condition. When the same average is performed in Eq.\eqref{gren1}, the result is $1+b_0/8$, with $b_{0}=2N/k_{0}^{2}R^{2}$ the cloud resonant optical thickness (see SM), i.e., the superradiant modes are largely populated. In the SM, we show that uncorrelated phases from the atomic positions bring the average decay rate back to the single atom one, which confirms that the superradiance stems from the phase coherence~\cite{wodkiewicz2006}.

Before we proceed to studying the radiated power, let us remind that the linear scaling with $b_0$ for super-~\cite{kaiser2011,guerin2016} and subradiance~\cite{kaiser2016,kaiser2017} has been observed in the linear optics regime, and this parameter has been identified as the control parameter for cooperative phenomena in dilute clouds. For the nonlinear regime, the first-order rate $\bar{\gamma}_z^{\left(l\right)}$ indicates that $b_0$ has to be corrected by a factor $1-p_e$, so we define $\mathcal{C}\equiv\left(1-p_e\right)b_0/8$ as the semi-classical cooperative parameter. Accordingly, solution \eqref{analy} must be valid in the regime $\mathcal{C} \ll 1$.

The first-order solution \eqref{analy} allows us to obtain a second-order approximation for the total power radiated (Eq.\eqref{P_nlin}). In the regime of vanishing interactions ($u_{lm}\rightarrow 0$, $\forall m\neq l$), the decay of independent atoms ($N p_e e^{-\Gamma t}$) is recovered, as expected. Interestingly, the same free decay is equally obtained when all dipoles tend to become fully-inverted $p_e\rightarrow 1$, an evidence that cooperation might be suppressed by atomic population ($\mathcal{C}\rightarrow 0$). Yet the complete absence of cooperation for $p_e=1$, which is observed both in numerical simulations and from the renormalization approach (see SM), is an incorrect prediction of the semi-classical truncation, as the emission of the first photons must be treated fully quantum-mechanically~\cite{arecchi1972,macgillivray1976}.

In Fig.\ref{fig2}(a), exact classical (linear optics) and semi-classical total powers are compared with the one predicted by the analytical solution \eqref{analy}, for half excited population and a cooperativity parameter $\mathcal{C}\approx0.04$. The linear-optics decay appears farther from the single-atom case than that calculated from the semi-classical model and fails at giving an accurate prediction for the semi-classical emission. However, the renormalization approach provides a very good approximation for the exact semi-classical emission over several $\Gamma t$, including the whole superradiant stage and a long period of the subradiant stage. Yet the superradiance effect is to small to be observed.
\begin{figure}[hbt]	
	\begin{center}
		\includegraphics[scale=0.42]{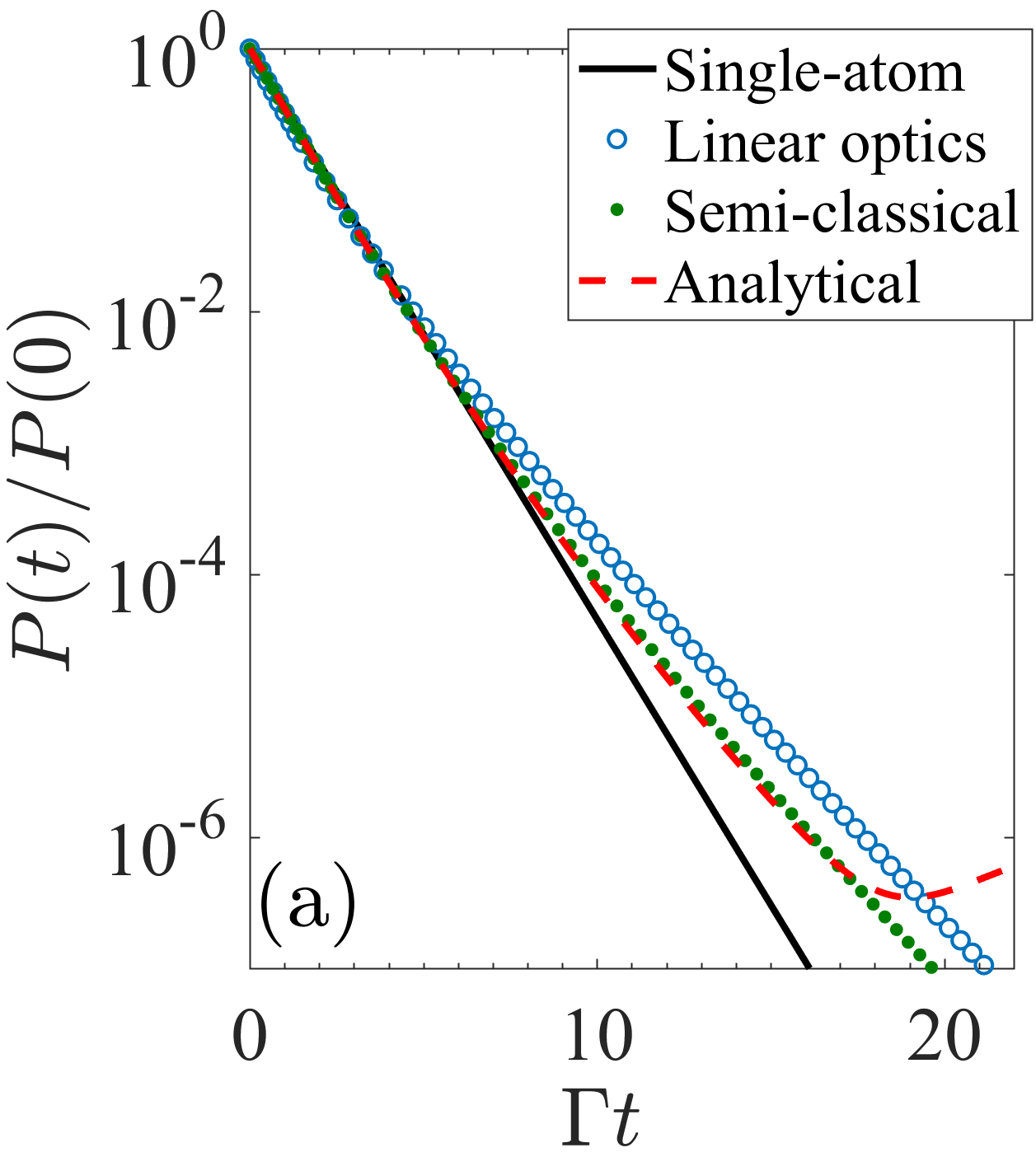}
		\includegraphics[scale=0.42]{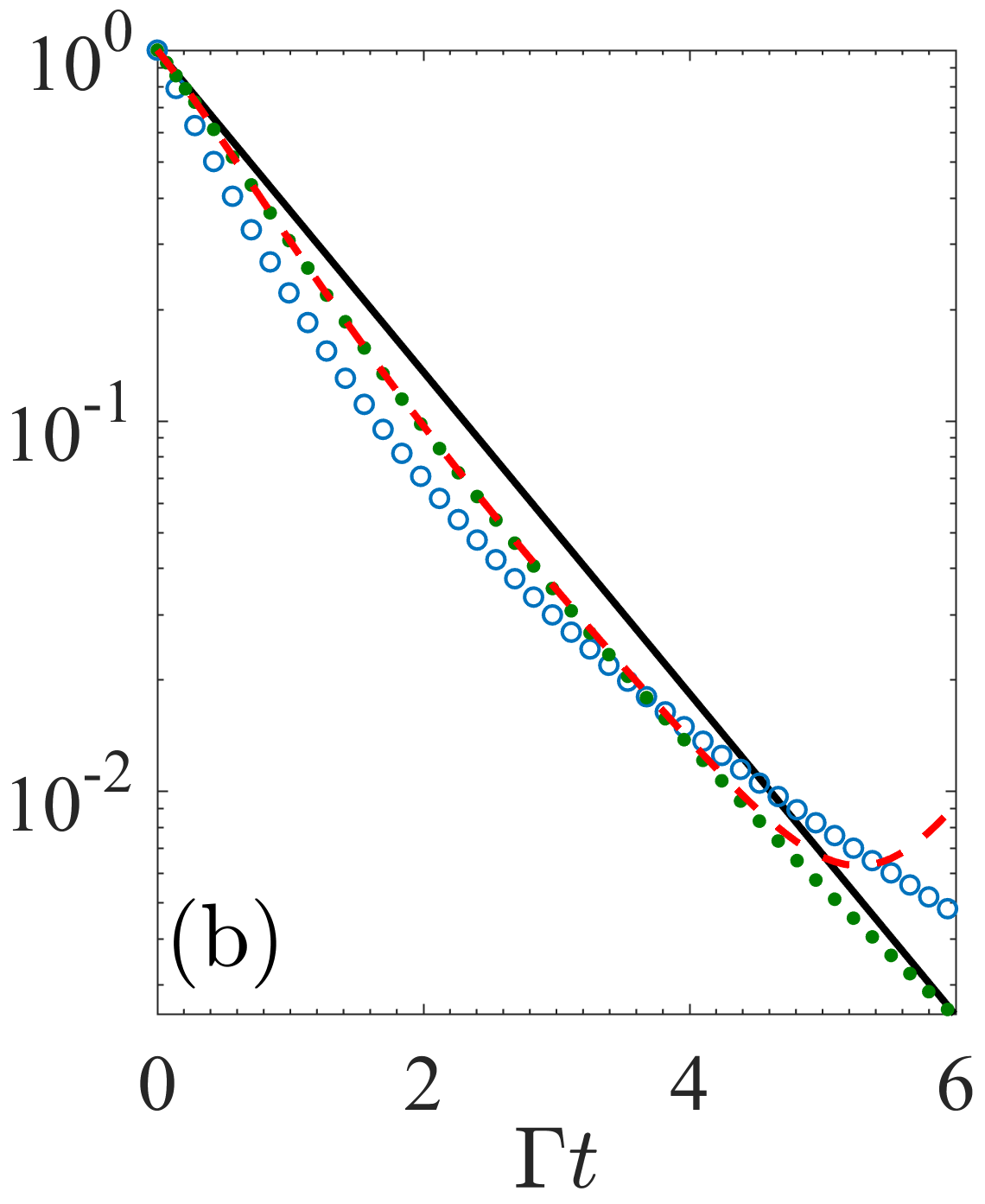}
	\end{center}
	\caption{\label{fig2} Normalized total power as a function of $\Gamma t$. The black solid curve stands for the single-atom case, the blue circles for the numerical linear-optics model, the green dots for the the numerical semi-classical model and the dashed red curve for the analytical solution. Simulations realized for a Gaussian cloud of $N=25000$ particles, (a) $R=45\lambda$ ($b_0\approx0.63$) and $p_e=0.5$ ($\mathcal{C} \approx 0.04$), (b) $R=20\lambda$ ($b_0\approx3.2$) and $p_e=0.6$ ($\mathcal{C} \approx 0.16$).}
\end{figure}

In order to clearly see superradiance, in Fig.\ref{fig2}(b) we have increased the cooperation to $\mathcal{C} \approx 0.16$. Note that the analytical solution is still capable to describe the semi-classical emission, which is expressively less accelerated than the classical case. However, the agreement with the numerical prediction ends up lasting for a smaller temporal range. Therefore, the parameter $\mathcal{C}$ also controls the temporal validity of solution \eqref{analy}: an accurate description of the dynamics is achieved for stronger cooperation, at the expense of a reduction of the trustful temporal range. It happens because, as the cooperation increases, the most subradiant rates becomes negative, predicting an unphysical energy gain in the emitted power.

To conclude, in this work we have proposed a renormalization formalism to solve truncated nonlinear equations of quantum many-body systems. The feasibility of our method was demonstrated by exploring cooperative spontaneous emission in the semi-classical regime, where a quadratic nonlinearity was addressed. However, the procedure can be straightforwardly extended to other higher-order cumulant expansions. The renormalization technique presents the advantage of solving these nonlinear problems for larger physical systems, with real possibilities of describing properly the thermodynamic limit (see SM). Different from numerical methods, we obtain analytical expressions for many-body dissipative rates and energy shifts, allowing for easy computational parallelization and more accurate physical interpretations.

As shown here at the lowest-order in the cumulant expansion, the interaction for renormalized solutions can be strong enough to distinguish higher-order quantum effects in connected correlations from lower-order ones. Hence collective effects of a truly quantum nature can be associated with a given order of the cumulant expansion for larger systems.  The approach is particularly promising to investigate non-equilibrium many-body physics in two and three dimensions, for short and long-range interacting systems.

\acknowledgments
We thank fruitful discussions with Robin Kaiser. C. E. M., R. B., F. E. A. S. and C. J. V. B. benefited from Grants from S\~ao Paulo Research Foundation (FAPESP) (Grant Nos. 2013/04162-5, 2014/01491-0, 2017/13250-6, 2018/01447-2 and 2013/07276-1) and from the National Council for Scientific and Technological Development (CNPq) Grant Nos.  305586/2017-3, 302981/2017-9, 409946/2018-4 and 307077/2018-7. The Titan X Pascal used for this research was donated by the NVIDIA Corporation. 

\bibliographystyle{apsrev4-1}
\bibliography{nlinear_ref}

\begin{thebibliography}{33}%
\makeatletter
\providecommand \@ifxundefined [1]{%
 \@ifx{#1\undefined}
}%
\providecommand \@ifnum [1]{%
 \ifnum #1\expandafter \@firstoftwo
 \else \expandafter \@secondoftwo
 \fi
}%
\providecommand \@ifx [1]{%
 \ifx #1\expandafter \@firstoftwo
 \else \expandafter \@secondoftwo
 \fi
}%
\providecommand \natexlab [1]{#1}%
\providecommand \enquote  [1]{``#1''}%
\providecommand \bibnamefont  [1]{#1}%
\providecommand \bibfnamefont [1]{#1}%
\providecommand \citenamefont [1]{#1}%
\providecommand \href@noop [0]{\@secondoftwo}%
\providecommand \href [0]{\begingroup \@sanitize@url \@href}%
\providecommand \@href[1]{\@@startlink{#1}\@@href}%
\providecommand \@@href[1]{\endgroup#1\@@endlink}%
\providecommand \@sanitize@url [0]{\catcode `\\12\catcode `\$12\catcode
  `\&12\catcode `\#12\catcode `\^12\catcode `\_12\catcode `\%12\relax}%
\providecommand \@@startlink[1]{}%
\providecommand \@@endlink[0]{}%
\providecommand \url  [0]{\begingroup\@sanitize@url \@url }%
\providecommand \@url [1]{\endgroup\@href {#1}{\urlprefix }}%
\providecommand \urlprefix  [0]{URL }%
\providecommand \Eprint [0]{\href }%
\providecommand \doibase [0]{http://dx.doi.org/}%
\providecommand \selectlanguage [0]{\@gobble}%
\providecommand \bibinfo  [0]{\@secondoftwo}%
\providecommand \bibfield  [0]{\@secondoftwo}%
\providecommand \translation [1]{[#1]}%
\providecommand \BibitemOpen [0]{}%
\providecommand \bibitemStop [0]{}%
\providecommand \bibitemNoStop [0]{.\EOS\space}%
\providecommand \EOS [0]{\spacefactor3000\relax}%
\providecommand \BibitemShut  [1]{\csname bibitem#1\endcsname}%
\let\auto@bib@innerbib\@empty
\bibitem [{\citenamefont {{Kr\"amer, Sebastian}}\ and\ \citenamefont {{Ritsch,
  Helmut}}(2015)}]{helmut2015}%
  \BibitemOpen
  \bibfield  {author} {\bibinfo {author} {\bibnamefont {{Kr\"amer,
  Sebastian}}}\ and\ \bibinfo {author} {\bibnamefont {{Ritsch, Helmut}}},\
  }\href {\doibase 10.1140/epjd/e2015-60266-5} {\bibfield  {journal} {\bibinfo
  {journal} {Eur. Phys. J. D}\ }\textbf {\bibinfo {volume} {69}},\ \bibinfo
  {pages} {282} (\bibinfo {year} {2015})}\BibitemShut {NoStop}%
\bibitem [{\citenamefont {Schachenmayer}\ \emph
  {et~al.}(2015{\natexlab{a}})\citenamefont {Schachenmayer}, \citenamefont
  {Pikovski},\ and\ \citenamefont {Rey}}]{1rey2015}%
  \BibitemOpen
  \bibfield  {author} {\bibinfo {author} {\bibfnamefont {J.}~\bibnamefont
  {Schachenmayer}}, \bibinfo {author} {\bibfnamefont {A.}~\bibnamefont
  {Pikovski}}, \ and\ \bibinfo {author} {\bibfnamefont {A.~M.}\ \bibnamefont
  {Rey}},\ }\href {\doibase 10.1103/PhysRevX.5.011022} {\bibfield  {journal}
  {\bibinfo  {journal} {Phys. Rev. X}\ }\textbf {\bibinfo {volume} {5}},\
  \bibinfo {pages} {011022} (\bibinfo {year} {2015}{\natexlab{a}})}\BibitemShut
  {NoStop}%
\bibitem [{\citenamefont {Schachenmayer}\ \emph
  {et~al.}(2015{\natexlab{b}})\citenamefont {Schachenmayer}, \citenamefont
  {Pikovski},\ and\ \citenamefont {Rey}}]{2rey2015}%
  \BibitemOpen
  \bibfield  {author} {\bibinfo {author} {\bibfnamefont {J.}~\bibnamefont
  {Schachenmayer}}, \bibinfo {author} {\bibfnamefont {A.}~\bibnamefont
  {Pikovski}}, \ and\ \bibinfo {author} {\bibfnamefont {A.~M.}\ \bibnamefont
  {Rey}},\ }\href {http://stacks.iop.org/1367-2630/17/i=6/a=065009} {\bibfield
  {journal} {\bibinfo  {journal} {New Journal of Physics}\ }\textbf {\bibinfo
  {volume} {17}},\ \bibinfo {pages} {065009} (\bibinfo {year}
  {2015}{\natexlab{b}})}\BibitemShut {NoStop}%
\bibitem [{\citenamefont {Pucci}\ \emph {et~al.}(2016)\citenamefont {Pucci},
  \citenamefont {Roy},\ and\ \citenamefont {Kastner}}]{kastner2016}%
  \BibitemOpen
  \bibfield  {author} {\bibinfo {author} {\bibfnamefont {L.}~\bibnamefont
  {Pucci}}, \bibinfo {author} {\bibfnamefont {A.}~\bibnamefont {Roy}}, \ and\
  \bibinfo {author} {\bibfnamefont {M.}~\bibnamefont {Kastner}},\ }\href
  {\doibase 10.1103/PhysRevB.93.174302} {\bibfield  {journal} {\bibinfo
  {journal} {Phys. Rev. B}\ }\textbf {\bibinfo {volume} {93}},\ \bibinfo
  {pages} {174302} (\bibinfo {year} {2016})}\BibitemShut {NoStop}%
\bibitem [{\citenamefont {Pucci}\ \emph {et~al.}(2017)\citenamefont {Pucci},
  \citenamefont {Roy}, \citenamefont {do~Espirito~Santo}, \citenamefont
  {Kaiser}, \citenamefont {Kastner},\ and\ \citenamefont
  {Bachelard}}]{bachelard2017}%
  \BibitemOpen
  \bibfield  {author} {\bibinfo {author} {\bibfnamefont {L.}~\bibnamefont
  {Pucci}}, \bibinfo {author} {\bibfnamefont {A.}~\bibnamefont {Roy}}, \bibinfo
  {author} {\bibfnamefont {T.~S.}\ \bibnamefont {do~Espirito~Santo}}, \bibinfo
  {author} {\bibfnamefont {R.}~\bibnamefont {Kaiser}}, \bibinfo {author}
  {\bibfnamefont {M.}~\bibnamefont {Kastner}}, \ and\ \bibinfo {author}
  {\bibfnamefont {R.}~\bibnamefont {Bachelard}},\ }\href {\doibase
  10.1103/PhysRevA.95.053625} {\bibfield  {journal} {\bibinfo  {journal} {Phys.
  Rev. A}\ }\textbf {\bibinfo {volume} {95}},\ \bibinfo {pages} {053625}
  (\bibinfo {year} {2017})}\BibitemShut {NoStop}%
\bibitem [{\citenamefont {Pi\~neiro Orioli}\ \emph {et~al.}(2017)\citenamefont
  {Pi\~neiro Orioli}, \citenamefont {Safavi-Naini}, \citenamefont {Wall},\ and\
  \citenamefont {Rey}}]{rey2017}%
  \BibitemOpen
  \bibfield  {author} {\bibinfo {author} {\bibfnamefont {A.}~\bibnamefont
  {Pi\~neiro Orioli}}, \bibinfo {author} {\bibfnamefont {A.}~\bibnamefont
  {Safavi-Naini}}, \bibinfo {author} {\bibfnamefont {M.~L.}\ \bibnamefont
  {Wall}}, \ and\ \bibinfo {author} {\bibfnamefont {A.~M.}\ \bibnamefont
  {Rey}},\ }\href {\doibase 10.1103/PhysRevA.96.033607} {\bibfield  {journal}
  {\bibinfo  {journal} {Phys. Rev. A}\ }\textbf {\bibinfo {volume} {96}},\
  \bibinfo {pages} {033607} (\bibinfo {year} {2017})}\BibitemShut {NoStop}%
\bibitem [{\citenamefont {Zhu}\ \emph {et~al.}(2016)\citenamefont {Zhu},
  \citenamefont {Cooper}, \citenamefont {Ye},\ and\ \citenamefont
  {Rey}}]{rey2016}%
  \BibitemOpen
  \bibfield  {author} {\bibinfo {author} {\bibfnamefont {B.}~\bibnamefont
  {Zhu}}, \bibinfo {author} {\bibfnamefont {J.}~\bibnamefont {Cooper}},
  \bibinfo {author} {\bibfnamefont {J.}~\bibnamefont {Ye}}, \ and\ \bibinfo
  {author} {\bibfnamefont {A.~M.}\ \bibnamefont {Rey}},\ }\href {\doibase
  10.1103/PhysRevA.94.023612} {\bibfield  {journal} {\bibinfo  {journal} {Phys.
  Rev. A}\ }\textbf {\bibinfo {volume} {94}},\ \bibinfo {pages} {023612}
  (\bibinfo {year} {2016})}\BibitemShut {NoStop}%
\bibitem [{\citenamefont {Lindstedt}(1883)}]{lindstedt1883}%
  \BibitemOpen
  \bibfield  {author} {\bibinfo {author} {\bibfnamefont {A.}~\bibnamefont
  {Lindstedt}},\ }\href {\doibase 10.1002/asna.18831041002} {\bibfield
  {journal} {\bibinfo  {journal} {Astronomische Nachrichten}\ }\textbf
  {\bibinfo {volume} {104}},\ \bibinfo {pages} {145} (\bibinfo {year}
  {1883})}\BibitemShut {NoStop}%
\bibitem [{\citenamefont {Bender}\ and\ \citenamefont
  {A.~Orszag}(1999)}]{book1}%
  \BibitemOpen
  \bibfield  {author} {\bibinfo {author} {\bibfnamefont {C.}~\bibnamefont
  {Bender}}\ and\ \bibinfo {author} {\bibfnamefont {S.}~\bibnamefont
  {A.~Orszag}}\ }(\bibinfo  {publisher} {Springer},\ \bibinfo {year}
  {1999})\BibitemShut {NoStop}%
\bibitem [{\citenamefont {Pelster}\ \emph {et~al.}(2003)\citenamefont
  {Pelster}, \citenamefont {Kleinert},\ and\ \citenamefont
  {Schanz}}]{schanz2003}%
  \BibitemOpen
  \bibfield  {author} {\bibinfo {author} {\bibfnamefont {A.}~\bibnamefont
  {Pelster}}, \bibinfo {author} {\bibfnamefont {H.}~\bibnamefont {Kleinert}}, \
  and\ \bibinfo {author} {\bibfnamefont {M.}~\bibnamefont {Schanz}},\ }\href
  {\doibase 10.1103/PhysRevE.67.016604} {\bibfield  {journal} {\bibinfo
  {journal} {Phys. Rev. E}\ }\textbf {\bibinfo {volume} {67}},\ \bibinfo
  {pages} {016604} (\bibinfo {year} {2003})}\BibitemShut {NoStop}%
\bibitem [{\citenamefont {Vidanovi\ifmmode~\acute{c}\else \'{c}\fi{}}\ \emph
  {et~al.}(2011)\citenamefont {Vidanovi\ifmmode~\acute{c}\else \'{c}\fi{}},
  \citenamefont {Bala\ifmmode~\check{z}\else \v{z}\fi{}}, \citenamefont
  {Al-Jibbouri},\ and\ \citenamefont {Pelster}}]{pelster2011}%
  \BibitemOpen
  \bibfield  {author} {\bibinfo {author} {\bibfnamefont {I.}~\bibnamefont
  {Vidanovi\ifmmode~\acute{c}\else \'{c}\fi{}}}, \bibinfo {author}
  {\bibfnamefont {A.}~\bibnamefont {Bala\ifmmode~\check{z}\else \v{z}\fi{}}},
  \bibinfo {author} {\bibfnamefont {H.}~\bibnamefont {Al-Jibbouri}}, \ and\
  \bibinfo {author} {\bibfnamefont {A.}~\bibnamefont {Pelster}},\ }\href
  {\doibase 10.1103/PhysRevA.84.013618} {\bibfield  {journal} {\bibinfo
  {journal} {Phys. Rev. A}\ }\textbf {\bibinfo {volume} {84}},\ \bibinfo
  {pages} {013618} (\bibinfo {year} {2011})}\BibitemShut {NoStop}%
\bibitem [{\citenamefont {Al-Jibbouri}\ \emph {et~al.}(2013)\citenamefont
  {Al-Jibbouri}, \citenamefont {Vidanovic}, \citenamefont {Balaz},\ and\
  \citenamefont {Pelster}}]{pelster2013}%
  \BibitemOpen
  \bibfield  {author} {\bibinfo {author} {\bibfnamefont {H.}~\bibnamefont
  {Al-Jibbouri}}, \bibinfo {author} {\bibfnamefont {I.}~\bibnamefont
  {Vidanovic}}, \bibinfo {author} {\bibfnamefont {A.}~\bibnamefont {Balaz}}, \
  and\ \bibinfo {author} {\bibfnamefont {A.}~\bibnamefont {Pelster}},\
  }\href@noop {} {\bibfield  {journal} {\bibinfo  {journal} {Journal of Physics
  B: Atomic, Molecular and Optical Physics}\ }\textbf {\bibinfo {volume}
  {46}},\ \bibinfo {pages} {065303} (\bibinfo {year} {2013})}\BibitemShut
  {NoStop}%
\bibitem [{\citenamefont {Arecchi}\ \emph {et~al.}(1972)\citenamefont
  {Arecchi}, \citenamefont {Courtens}, \citenamefont {Gilmore},\ and\
  \citenamefont {Thomas}}]{arecchi1972}%
  \BibitemOpen
  \bibfield  {author} {\bibinfo {author} {\bibfnamefont {F.~T.}\ \bibnamefont
  {Arecchi}}, \bibinfo {author} {\bibfnamefont {E.}~\bibnamefont {Courtens}},
  \bibinfo {author} {\bibfnamefont {R.}~\bibnamefont {Gilmore}}, \ and\
  \bibinfo {author} {\bibfnamefont {H.}~\bibnamefont {Thomas}},\ }\href
  {\doibase 10.1103/physreva.6.2211} {\bibfield  {journal} {\bibinfo  {journal}
  {Physical Review A}\ }\textbf {\bibinfo {volume} {6}},\ \bibinfo {pages}
  {2211} (\bibinfo {year} {1972})}\BibitemShut {NoStop}%
\bibitem [{\citenamefont {MacGillivray}\ and\ \citenamefont
  {Feld}(1976)}]{macgillivray1976}%
  \BibitemOpen
  \bibfield  {author} {\bibinfo {author} {\bibfnamefont {J.~C.}\ \bibnamefont
  {MacGillivray}}\ and\ \bibinfo {author} {\bibfnamefont {M.~S.}\ \bibnamefont
  {Feld}},\ }\href {\doibase 10.1103/physreva.14.1169} {\bibfield  {journal}
  {\bibinfo  {journal} {Physical Review A}\ }\textbf {\bibinfo {volume} {14}},\
  \bibinfo {pages} {1169} (\bibinfo {year} {1976})}\BibitemShut {NoStop}%
\bibitem [{\citenamefont {Scully}\ \emph {et~al.}(2006)\citenamefont {Scully},
  \citenamefont {Fry}, \citenamefont {Ooi},\ and\ \citenamefont
  {W\'odkiewicz}}]{wodkiewicz2006}%
  \BibitemOpen
  \bibfield  {author} {\bibinfo {author} {\bibfnamefont {M.~O.}\ \bibnamefont
  {Scully}}, \bibinfo {author} {\bibfnamefont {E.~S.}\ \bibnamefont {Fry}},
  \bibinfo {author} {\bibfnamefont {C.~H.~R.}\ \bibnamefont {Ooi}}, \ and\
  \bibinfo {author} {\bibfnamefont {K.}~\bibnamefont {W\'odkiewicz}},\ }\href
  {\doibase 10.1103/PhysRevLett.96.010501} {\bibfield  {journal} {\bibinfo
  {journal} {Physical Review Letters}\ }\textbf {\bibinfo {volume} {96}},\
  \bibinfo {pages} {010501} (\bibinfo {year} {2006})}\BibitemShut {NoStop}%
\bibitem [{\citenamefont {Mazets}\ and\ \citenamefont
  {Kurizki}(2007)}]{kurizki2007}%
  \BibitemOpen
  \bibfield  {author} {\bibinfo {author} {\bibfnamefont {I.~E.}\ \bibnamefont
  {Mazets}}\ and\ \bibinfo {author} {\bibfnamefont {G.}~\bibnamefont
  {Kurizki}},\ }\href@noop {} {\bibfield  {journal} {\bibinfo  {journal}
  {Journal of Physics B}\ }\textbf {\bibinfo {volume} {40}},\ \bibinfo {pages}
  {F105} (\bibinfo {year} {2007})}\BibitemShut {NoStop}%
\bibitem [{\citenamefont {Svidzinsky}\ \emph {et~al.}(2008)\citenamefont
  {Svidzinsky}, \citenamefont {Chang},\ and\ \citenamefont
  {Scully}}]{1scully2008}%
  \BibitemOpen
  \bibfield  {author} {\bibinfo {author} {\bibfnamefont {A.~A.}\ \bibnamefont
  {Svidzinsky}}, \bibinfo {author} {\bibfnamefont {J.-T.}\ \bibnamefont
  {Chang}}, \ and\ \bibinfo {author} {\bibfnamefont {M.~O.}\ \bibnamefont
  {Scully}},\ }\href {\doibase 10.1103/PhysRevLett.100.160504} {\bibfield
  {journal} {\bibinfo  {journal} {Physical Review Letters}\ }\textbf {\bibinfo
  {volume} {100}},\ \bibinfo {pages} {160504} (\bibinfo {year}
  {2008})}\BibitemShut {NoStop}%
\bibitem [{\citenamefont {Bux}\ \emph {et~al.}(2010)\citenamefont {Bux},
  \citenamefont {Lucioni}, \citenamefont {Bender}, \citenamefont {Bienaim\'e},
  \citenamefont {Lauber}, \citenamefont {Stehle}, \citenamefont {Zimmermann},
  \citenamefont {Slama}, \citenamefont {Courteille}, \citenamefont {Piovella},\
  and\ \citenamefont {Kaiser}}]{1kaiser2010}%
  \BibitemOpen
  \bibfield  {author} {\bibinfo {author} {\bibfnamefont {S.}~\bibnamefont
  {Bux}}, \bibinfo {author} {\bibfnamefont {E.}~\bibnamefont {Lucioni}},
  \bibinfo {author} {\bibfnamefont {H.}~\bibnamefont {Bender}}, \bibinfo
  {author} {\bibfnamefont {T.}~\bibnamefont {Bienaim\'e}}, \bibinfo {author}
  {\bibfnamefont {K.}~\bibnamefont {Lauber}}, \bibinfo {author} {\bibfnamefont
  {C.}~\bibnamefont {Stehle}}, \bibinfo {author} {\bibfnamefont
  {C.}~\bibnamefont {Zimmermann}}, \bibinfo {author} {\bibfnamefont
  {S.}~\bibnamefont {Slama}}, \bibinfo {author} {\bibfnamefont
  {P.}~\bibnamefont {Courteille}}, \bibinfo {author} {\bibfnamefont
  {N.}~\bibnamefont {Piovella}}, \ and\ \bibinfo {author} {\bibfnamefont
  {R.}~\bibnamefont {Kaiser}},\ }\href {\doibase 10.1080/09500340.2010.503011}
  {\bibfield  {journal} {\bibinfo  {journal} {Journal of Modern Optics}\
  }\textbf {\bibinfo {volume} {57}},\ \bibinfo {pages} {1841} (\bibinfo {year}
  {2010})}\BibitemShut {NoStop}%
\bibitem [{\citenamefont {Svidzinsky}\ \emph {et~al.}(2010)\citenamefont
  {Svidzinsky}, \citenamefont {Chang},\ and\ \citenamefont
  {Scully}}]{scully2010}%
  \BibitemOpen
  \bibfield  {author} {\bibinfo {author} {\bibfnamefont {A.~A.}\ \bibnamefont
  {Svidzinsky}}, \bibinfo {author} {\bibfnamefont {J.-T.}\ \bibnamefont
  {Chang}}, \ and\ \bibinfo {author} {\bibfnamefont {M.~O.}\ \bibnamefont
  {Scully}},\ }\href {\doibase 10.1103/PhysRevA.81.053821} {\bibfield
  {journal} {\bibinfo  {journal} {Physical Review A}\ }\textbf {\bibinfo
  {volume} {81}},\ \bibinfo {pages} {053821} (\bibinfo {year}
  {2010})}\BibitemShut {NoStop}%
\bibitem [{\citenamefont {Bienaim\'e}\ \emph {et~al.}(2011)\citenamefont
  {Bienaim\'e}, \citenamefont {Petruzzo}, \citenamefont {Bigerni},
  \citenamefont {Piovella},\ and\ \citenamefont {Kaiser}}]{kaiser2011}%
  \BibitemOpen
  \bibfield  {author} {\bibinfo {author} {\bibfnamefont {T.}~\bibnamefont
  {Bienaim\'e}}, \bibinfo {author} {\bibfnamefont {M.}~\bibnamefont
  {Petruzzo}}, \bibinfo {author} {\bibfnamefont {D.}~\bibnamefont {Bigerni}},
  \bibinfo {author} {\bibfnamefont {N.}~\bibnamefont {Piovella}}, \ and\
  \bibinfo {author} {\bibfnamefont {R.}~\bibnamefont {Kaiser}},\ }\href
  {\doibase 10.1080/09500340.2011.594911} {\bibfield  {journal} {\bibinfo
  {journal} {Journal of Modern Optics}\ }\textbf {\bibinfo {volume} {58}},\
  \bibinfo {pages} {1942} (\bibinfo {year} {2011})}\BibitemShut {NoStop}%
\bibitem [{\citenamefont {Bienaim\'e}\ \emph {et~al.}(2013)\citenamefont
  {Bienaim\'e}, \citenamefont {Bachelard}, \citenamefont {Piovella},\ and\
  \citenamefont {Kaiser}}]{kaiser2012}%
  \BibitemOpen
  \bibfield  {author} {\bibinfo {author} {\bibfnamefont {T.}~\bibnamefont
  {Bienaim\'e}}, \bibinfo {author} {\bibfnamefont {R.}~\bibnamefont
  {Bachelard}}, \bibinfo {author} {\bibfnamefont {N.}~\bibnamefont {Piovella}},
  \ and\ \bibinfo {author} {\bibfnamefont {R.}~\bibnamefont {Kaiser}},\ }\href
  {\doibase 10.1002/prop.201200089} {\bibfield  {journal} {\bibinfo  {journal}
  {Fortschritte der Physik}\ }\textbf {\bibinfo {volume} {61}},\ \bibinfo
  {pages} {377} (\bibinfo {year} {2013})}\BibitemShut {NoStop}%
\bibitem [{\citenamefont {Svidzinsky}\ \emph {et~al.}(2015)\citenamefont
  {Svidzinsky}, \citenamefont {Zhang},\ and\ \citenamefont
  {Scully}}]{scully2015}%
  \BibitemOpen
  \bibfield  {author} {\bibinfo {author} {\bibfnamefont {A.~A.}\ \bibnamefont
  {Svidzinsky}}, \bibinfo {author} {\bibfnamefont {X.}~\bibnamefont {Zhang}}, \
  and\ \bibinfo {author} {\bibfnamefont {M.~O.}\ \bibnamefont {Scully}},\
  }\href {\doibase 10.1103/PhysRevA.92.013801} {\bibfield  {journal} {\bibinfo
  {journal} {Physical Review A}\ }\textbf {\bibinfo {volume} {92}},\ \bibinfo
  {pages} {013801} (\bibinfo {year} {2015})}\BibitemShut {NoStop}%
\bibitem [{\citenamefont {Cottier}\ \emph {et~al.}(2018)\citenamefont
  {Cottier}, \citenamefont {Kaiser},\ and\ \citenamefont
  {Bachelard}}]{cottier2018}%
  \BibitemOpen
  \bibfield  {author} {\bibinfo {author} {\bibfnamefont {F.}~\bibnamefont
  {Cottier}}, \bibinfo {author} {\bibfnamefont {R.}~\bibnamefont {Kaiser}}, \
  and\ \bibinfo {author} {\bibfnamefont {R.}~\bibnamefont {Bachelard}},\ }\href
  {\doibase 10.1103/physreva.98.013622} {\bibfield  {journal} {\bibinfo
  {journal} {Physical Review A}\ }\textbf {\bibinfo {volume} {98}} (\bibinfo
  {year} {2018}),\ 10.1103/physreva.98.013622}\BibitemShut {NoStop}%
\bibitem [{\citenamefont {de~Oliveira}\ \emph {et~al.}(2014)\citenamefont
  {de~Oliveira}, \citenamefont {Mendes}, \citenamefont {Martins}, \citenamefont
  {Saldanha}, \citenamefont {Tabosa},\ and\ \citenamefont
  {Felinto}}]{felinto2014}%
  \BibitemOpen
  \bibfield  {author} {\bibinfo {author} {\bibfnamefont {R.~A.}\ \bibnamefont
  {de~Oliveira}}, \bibinfo {author} {\bibfnamefont {M.~S.}\ \bibnamefont
  {Mendes}}, \bibinfo {author} {\bibfnamefont {W.~S.}\ \bibnamefont {Martins}},
  \bibinfo {author} {\bibfnamefont {P.~L.}\ \bibnamefont {Saldanha}}, \bibinfo
  {author} {\bibfnamefont {J.~W.~R.}\ \bibnamefont {Tabosa}}, \ and\ \bibinfo
  {author} {\bibfnamefont {D.}~\bibnamefont {Felinto}},\ }\href {\doibase
  10.1103/PhysRevA.90.023848} {\bibfield  {journal} {\bibinfo  {journal}
  {Physical Review A}\ }\textbf {\bibinfo {volume} {90}},\ \bibinfo {pages}
  {023848} (\bibinfo {year} {2014})}\BibitemShut {NoStop}%
\bibitem [{\citenamefont {Ara{\'u}jo}\ \emph {et~al.}(2016)\citenamefont
  {Ara{\'u}jo}, \citenamefont {Kre{\v{s}}i{\'c}}, \citenamefont {Kaiser},\ and\
  \citenamefont {Guerin}}]{guerin2016}%
  \BibitemOpen
  \bibfield  {author} {\bibinfo {author} {\bibfnamefont {M.~O.}\ \bibnamefont
  {Ara{\'u}jo}}, \bibinfo {author} {\bibfnamefont {I.}~\bibnamefont
  {Kre{\v{s}}i{\'c}}}, \bibinfo {author} {\bibfnamefont {R.}~\bibnamefont
  {Kaiser}}, \ and\ \bibinfo {author} {\bibfnamefont {W.}~\bibnamefont
  {Guerin}},\ }\href@noop {} {\bibfield  {journal} {\bibinfo  {journal}
  {Physical Review Letters}\ }\textbf {\bibinfo {volume} {117}},\ \bibinfo
  {pages} {073002} (\bibinfo {year} {2016})}\BibitemShut {NoStop}%
\bibitem [{\citenamefont {Roof}\ \emph {et~al.}(2016)\citenamefont {Roof},
  \citenamefont {Kemp}, \citenamefont {Havey},\ and\ \citenamefont
  {Sokolov}}]{roof2016}%
  \BibitemOpen
  \bibfield  {author} {\bibinfo {author} {\bibfnamefont {S.}~\bibnamefont
  {Roof}}, \bibinfo {author} {\bibfnamefont {K.}~\bibnamefont {Kemp}}, \bibinfo
  {author} {\bibfnamefont {M.}~\bibnamefont {Havey}}, \ and\ \bibinfo {author}
  {\bibfnamefont {I.}~\bibnamefont {Sokolov}},\ }\href {\doibase
  10.1103/physrevlett.117.073003} {\bibfield  {journal} {\bibinfo  {journal}
  {Physical Review Letters}\ }\textbf {\bibinfo {volume} {117}} (\bibinfo
  {year} {2016}),\ 10.1103/physrevlett.117.073003}\BibitemShut {NoStop}%
\bibitem [{\citenamefont {Eberly}(2006)}]{eberly2006}%
  \BibitemOpen
  \bibfield  {author} {\bibinfo {author} {\bibfnamefont {J.~H.}\ \bibnamefont
  {Eberly}},\ }\href@noop {} {\bibfield  {journal} {\bibinfo  {journal}
  {Journal of Physics B}\ }\textbf {\bibinfo {volume} {39}},\ \bibinfo {pages}
  {S599} (\bibinfo {year} {2006})}\BibitemShut {NoStop}%
\bibitem [{\citenamefont {Bienaim\'e}\ \emph {et~al.}(2012)\citenamefont
  {Bienaim\'e}, \citenamefont {Piovella},\ and\ \citenamefont
  {Kaiser}}]{1kaiser2012}%
  \BibitemOpen
  \bibfield  {author} {\bibinfo {author} {\bibfnamefont {T.}~\bibnamefont
  {Bienaim\'e}}, \bibinfo {author} {\bibfnamefont {N.}~\bibnamefont
  {Piovella}}, \ and\ \bibinfo {author} {\bibfnamefont {R.}~\bibnamefont
  {Kaiser}},\ }\href {\doibase 10.1103/PhysRevLett.108.123602} {\bibfield
  {journal} {\bibinfo  {journal} {Physical Review Letters}\ }\textbf {\bibinfo
  {volume} {108}},\ \bibinfo {pages} {123602} (\bibinfo {year}
  {2012})}\BibitemShut {NoStop}%
\bibitem [{\citenamefont {Guerin}\ \emph {et~al.}(2016)\citenamefont {Guerin},
  \citenamefont {Ara\'ujo},\ and\ \citenamefont {Kaiser}}]{kaiser2016}%
  \BibitemOpen
  \bibfield  {author} {\bibinfo {author} {\bibfnamefont {W.}~\bibnamefont
  {Guerin}}, \bibinfo {author} {\bibfnamefont {M.~O.}\ \bibnamefont
  {Ara\'ujo}}, \ and\ \bibinfo {author} {\bibfnamefont {R.}~\bibnamefont
  {Kaiser}},\ }\href {\doibase 10.1103/PhysRevLett.116.083601} {\bibfield
  {journal} {\bibinfo  {journal} {Physical Review Letters}\ }\textbf {\bibinfo
  {volume} {116}},\ \bibinfo {pages} {083601} (\bibinfo {year}
  {2016})}\BibitemShut {NoStop}%
\bibitem [{\citenamefont {Kr\"{a}mer}\ and\ \citenamefont
  {Ritsch}(2015)}]{Kramer2015}%
  \BibitemOpen
  \bibfield  {author} {\bibinfo {author} {\bibfnamefont {S.}~\bibnamefont
  {Kr\"{a}mer}}\ and\ \bibinfo {author} {\bibfnamefont {H.}~\bibnamefont
  {Ritsch}},\ }\href {\doibase 10.1140/epjd/e2015-60266-5} {\bibfield
  {journal} {\bibinfo  {journal} {The European Physical Journal D}\ }\textbf
  {\bibinfo {volume} {69}} (\bibinfo {year} {2015}),\
  10.1140/epjd/e2015-60266-5}\BibitemShut {NoStop}%
\bibitem [{\citenamefont {Young}\ \emph {et~al.}(2018)\citenamefont {Young},
  \citenamefont {Boulier}, \citenamefont {Magnan}, \citenamefont {Goldschmidt},
  \citenamefont {Wilson}, \citenamefont {Rolston}, \citenamefont {Porto},\ and\
  \citenamefont {Gorshkov}}]{Young2018}%
  \BibitemOpen
  \bibfield  {author} {\bibinfo {author} {\bibfnamefont {J.~T.}\ \bibnamefont
  {Young}}, \bibinfo {author} {\bibfnamefont {T.}~\bibnamefont {Boulier}},
  \bibinfo {author} {\bibfnamefont {E.}~\bibnamefont {Magnan}}, \bibinfo
  {author} {\bibfnamefont {E.~A.}\ \bibnamefont {Goldschmidt}}, \bibinfo
  {author} {\bibfnamefont {R.~M.}\ \bibnamefont {Wilson}}, \bibinfo {author}
  {\bibfnamefont {S.~L.}\ \bibnamefont {Rolston}}, \bibinfo {author}
  {\bibfnamefont {J.~V.}\ \bibnamefont {Porto}}, \ and\ \bibinfo {author}
  {\bibfnamefont {A.~V.}\ \bibnamefont {Gorshkov}},\ }\href {\doibase
  10.1103/physreva.97.023424} {\bibfield  {journal} {\bibinfo  {journal}
  {Physical Review A}\ }\textbf {\bibinfo {volume} {97}} (\bibinfo {year}
  {2018}),\ 10.1103/physreva.97.023424}\BibitemShut {NoStop}%
\bibitem [{\citenamefont {Guerin}\ \emph {et~al.}(2017)\citenamefont {Guerin},
  \citenamefont {Rouabah},\ and\ \citenamefont {Kaiser}}]{kaiser2017}%
  \BibitemOpen
  \bibfield  {author} {\bibinfo {author} {\bibfnamefont {W.}~\bibnamefont
  {Guerin}}, \bibinfo {author} {\bibfnamefont {M.}~\bibnamefont {Rouabah}}, \
  and\ \bibinfo {author} {\bibfnamefont {R.}~\bibnamefont {Kaiser}},\ }\href
  {\doibase 10.1080/09500340.2016.1215564} {\bibfield  {journal} {\bibinfo
  {journal} {Journal of Modern Optics}\ }\textbf {\bibinfo {volume} {64}},\
  \bibinfo {pages} {895} (\bibinfo {year} {2017})}\BibitemShut {NoStop}%
\bibitem [{\citenamefont {Scully}(2007)}]{scully2007}%
  \BibitemOpen
  \bibfield  {author} {\bibinfo {author} {\bibfnamefont {M.~O.}\ \bibnamefont
  {Scully}},\ }\href {\doibase 10.1134/S1054660X07050064} {\bibfield  {journal}
  {\bibinfo  {journal} {Laser Physics}\ }\textbf {\bibinfo {volume} {17}},\
  \bibinfo {pages} {635} (\bibinfo {year} {2007})}\BibitemShut {NoStop}%
\end{thebibliography}%

\onecolumngrid

\section{Supplementary Material}

\subsection{Green's function proof}
\renewcommand{\theequation}{S-\arabic{equation}}
  \setcounter{equation}{0}  
  
In this section, we prove that solution (\ref{G}) satisfies the Green's differential equation (\ref{dG}). After substituting expression (\ref{G}) in Eq.(\ref{dG}), we can use the eigenvalue equation $M_{ln}\psi_{n}^{\left(\alpha\right)}=\gamma^{\left(\alpha\right)}\psi_{l}^{\left(\alpha\right)}$ to obtain the condition
\begin{equation}
\delta\left(t-t^{\prime}\right)\left(\delta_{lm}-\sum_{\alpha}\psi_{l}^{\left(\alpha\right)}\phi_{m}^{\left(\alpha\right)}e^{-\gamma^{\left(\alpha\right)}\left(t-t^{\prime}\right)}\right)=0.\label{cond1}
\end{equation}
In the step above, it was assumed that the derivative of the Heaveside function results in the Dirac delta function. Note that condition Eq.(\ref{cond1}) is always true for $t\neq t^{\prime}$. For $t=t^{\prime}$, we obtain the following generalized closure relation:
\begin{equation}
\sum_{\alpha}\psi_{l}^{\left(\alpha\right)}\phi_{m}^{\left(\alpha\right)}=\delta_{lm}.\label{cond2}
\end{equation}

Multiplying both sides of Eq.(\ref{cond2}) by $\psi_{m}^{\left(\gamma\right)}$, we end up with
\begin{equation}
\sum_{\alpha}\psi_{l}^{\left(\alpha\right)}\phi_{m}^{\left(\alpha\right)}\psi_{m}^{\left(\gamma\right)}=\delta_{lm}\psi_{m}^{\left(\gamma\right)}.\label{cond3}
\end{equation}
Using the definition (\ref{phi}) in (\ref{cond3}), we finally prove our goal:
\begin{eqnarray}
\sum_{\alpha,\beta}\psi_{l}^{\left(\alpha\right)}\left[\mathbf{J}^{-1}\right]^{\left(\alpha\beta\right)}\psi_{m}^{\left(\beta\right)}\psi_{m}^{\left(\gamma\right)}&=&\psi_{l}^{\left(\gamma\right)},\\\sum_{\alpha,\beta}\psi_{l}^{\left(\alpha\right)}\left[\mathbf{J}^{-1}\right]^{\left(\alpha\beta\right)}\left[\mathbf{J}\right]^{\left(\beta\gamma\right)}&=&\psi_{l}^{\left(\gamma\right)},\\\sum_{\alpha}\psi_{l}^{\left(\alpha\right)}\delta_{\alpha\gamma}&=&\psi_{l}^{\left(\gamma\right)},\\\psi_{l}^{\left(\gamma\right)}&=&\psi_{l}^{\left(\gamma\right)}.
\end{eqnarray}

\subsection{Alternative way to derive the parameter $\mathcal{C}$ in the large $N$ limit}

It is possible to derive $\mathcal{C}$ from the total power radiated Eq.\eqref{P_nlin} at $t=0$, by evaluating the magnitude of the interacting term  of over the corresponding single atom term:
\begin{equation}
C\equiv2\frac{\left|\sum_{m=1}^{N}\sum_{n\neq m}\Re\left(u_{mn}\bigl\langle\hat{\sigma}_{m}^{+}\left(0\right)\bigr\rangle\bigl\langle\hat{\sigma}_{n}^{-}\left(0\right)\bigr\rangle\right)\right|}{\left|\sum_{m=1}^{N}\left(1+\bigl\langle\hat{\sigma}_{m}^{z}\left(0\right)\bigr\rangle\right)\right|}.
\end{equation}
Therefore, the light-mediated cooperation become substantial when $\mathcal{C}\gtrsim 1$. Considering the initial conditions \eqref{ini}, we obtain $\mathcal{C}=\left(1-p_{e}\right)\left|\Re\left(\left\langle \eta\right\rangle \right)\right|$, with $\left\langle \eta\right\rangle =\sum_{m=1}^{N}\eta_{m}/N$ the average value of the microscopic factor $\eta_m$ and $1-p_e$ the ground state population.

For macroscopic atomic clouds, with large number of atoms ($N\gg1$), we can calculate $\left\langle \eta\right\rangle$ substituting sums by integrals, namely,
\begin{equation}
\left\langle \eta\right\rangle =\frac{1}{N}\int d\mathbf{r}_{1}\rho\left(\mathbf{r}_{1}\right)\int d\mathbf{r}_{2}\rho\left(\mathbf{r}_{2}\right)\frac{\exp\left[ik_{0}\bigl|\mathbf{r}_{1}-\mathbf{r}_{2}\bigr|+i\mathbf{k}_{0}\cdot\left(\mathbf{r}_{1}-\mathbf{r}_{2}\right)\right]}{ik_{0}\bigl|\mathbf{r}_{1}-\mathbf{r}_{2}\bigr|},
\end{equation}
with $\rho\left(\mathbf{r}\right)$ the density distribution of the cloud. The change of variables, $ \mathbf{r}=\mathbf{r}_{1}-\mathbf{r}_{2}$ and $\mathbf{s}=\left(\mathbf{r}_{1}+\mathbf{r}_{2}\right)/2$, yields
\begin{equation}
\left\langle \eta\right\rangle =\frac{1}{N}\int d\mathbf{s}\int d\mathbf{r}\rho\left(\mathbf{s}-\frac{\mathbf{r}}{2}\right)\rho\left(\mathbf{s}+\frac{\mathbf{r}}{2}\right)\frac{\exp\left[ik_{0}\bigl|\mathbf{r}\bigr|+i\mathbf{k}_{0}\cdot\mathbf{r}\right]}{ik_{0}\bigl|\mathbf{r}\bigr|}.
\end{equation}
For a Gaussian distribution $\rho\left(\mathbf{r}\right)=\rho_0 \exp{\left(-r^{2}/2R^{2}\right)}$, with standard deviation $R$ and peak $\rho_0 \equiv N/ R^3 \left(2\pi\right)^{3/2}$, note that the following property holds:
\begin{equation}
\left|\mathbf{s}-\frac{\mathbf{r}}{2}\right|^{2}+\left|\mathbf{s}+\frac{\mathbf{r}}{2}\right|^{2}=2s^{2}+\frac{r^{2}}{2},
\end{equation}
so the radial integrals completely decouple from each other.

Considering that the incident plane wave propagates over the $z$ direction, we can express the phase of the exponential as $\mathbf{k}_{0}\cdot\mathbf{r}= k_{0}r\cos\theta$ to trivially solve the angular integral. In the limit of large atomic clouds ($k_{0}R\gg2\pi$), the remaining radial integrals
\begin{eqnarray}
\int_{0}^{\infty}e^{-\frac{s^{2}}{R^{2}}}s^{2}ds&=&\frac{\sqrt{\pi}}{4}R^{3},\\\int_{0}^{\infty}dre^{-\frac{r^{2}}{4R^{2}}}e^{ik_{0}r}\sin k_{0}r&\approx &i\frac{\sqrt{\pi}}{2}R,
\end{eqnarray}
results in $\left\langle \eta\right\rangle \approx b_0/8$ and $\mathcal{C}\approx(1-p_e)b_0/8$, with $b_0=2N/k_0^2R^2$ the cloud resonant optical thickness.

\subsection{Further discussions}

In the following, we demonstrate numerically that the perturbation parameter $\mathcal{C}$ is consistent with the cooperation reduction also for the exact models. We highlight that, we have detected no relevant differences between random realizations of the total power radiated for $N \geqslant 20000$. Also, in order to avoid significant contributions of atomic pairs, an exclusion volume of $0.3\lambda$ was assumed for all atoms in the entire work. 

\begin{figure}[htb]	
	\begin{center}
		\includegraphics[scale=0.42]{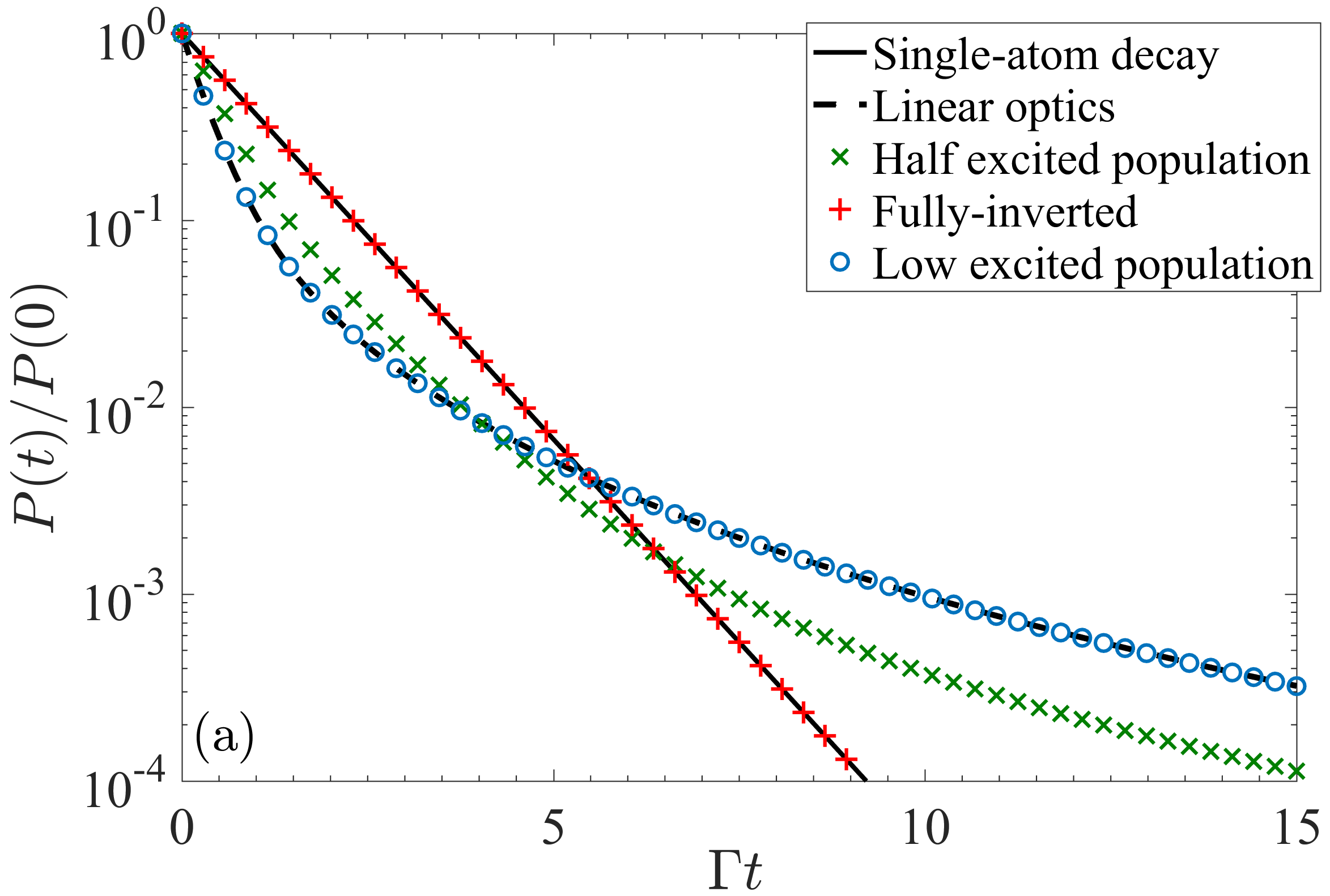}
		\includegraphics[scale=0.43]{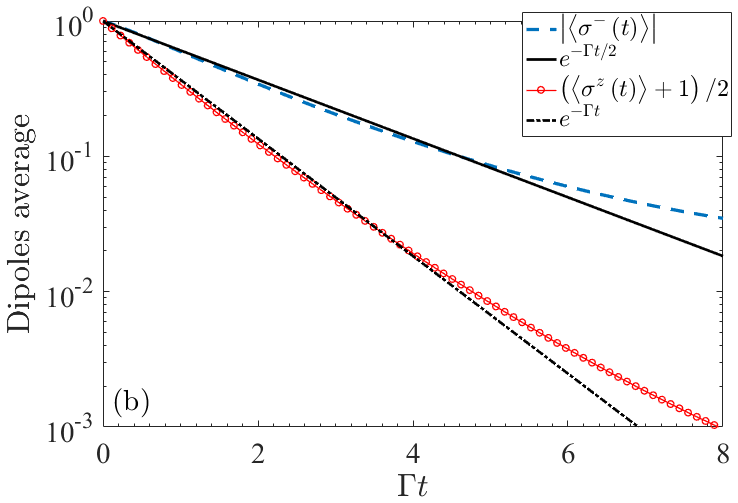}
	\end{center}
	\caption{\label{fig1_supp} (a) Normalized total power radiated for classical and semi-classical exact models, with the system decaying from the initial condition \eqref{ini}. The single-atom decay stands for the black solid curve and the linear optics decay stands for the black dashed curve. The symbols in color depict the semi-classical decay for a weakly excited system ($p_e=10^{-3}$, $\mathcal{C}\approx1.1$), half-excited system ($p_e=0.5$, $\mathcal{C}\approx0.55$) and fully-inverted system ($p_e=1$, $\mathcal{C}=0$). Simulations realized with $N=25000$ and $R=12\lambda$, which corresponds to a resonant optical thickness $b_0\approx8.8$. (b) Analytical dipole dynamics averaged over the whole atomic cloud. Simulations realized with half-excited system ($p_e=0.5$), for $N=10^5$ and $R=55\lambda$, which corresponds to $\mathcal{C}\approx0.1$ and $b_0\approx1.7$.} 
\end{figure}

In Fig.\ref{fig1_supp}(a), we present the normalized total power radiated, extracted from exact classical (linear optics) and semi-classical (nonlinear) models, where a superradiant stage is followed by a subradiant one. When the excited population is very low ($p_e \approx 10^{-3}$), note that $\mathcal{C} \approx b_{0}/8$ and the nonlinear model presents a behavior in excellent agreement with the linear optics model. However, as the excited population increases, a cooperativity reduction of $\mathcal{C}<b_{0}/8$ is expected for the nonlinear model as compared to the linear case. Indeed, the classical curve does not exhibit any change for $p_e = 0.5$ (not shown since all curves overlap exactly with the one at low $p_e$), whereas the semi-classical radiation become closer to the single-atom curve. In this regime, the excited population plays an important role on the coherence dynamics, which is not accounted by the linear optics model.

In the extreme case of a fully inverted system ($ p_e=1$), a complete absence of cooperation is expected ($\mathcal{C}=0$), as a consequence of the coherence cancelling in the Bloch state \eqref{ini}. In agreement with this prediction, we verify in Fig.\ref{fig1_supp}(a) that the numerical curve for the nonlinear model coincides with the single atom emission. In the superradiant cascade, the necessity of treating the emission of the first photons totally quantum-mechanically, until the vacuum modes are significantly populated and the semi-classical modelling can take over, has been pointed out in Ref.\cite{arecchi1972,macgillivray1976}. The present analysis thus confirms that the semi-classical approach does not describe properly that limit. Our analytical solution is naturally loyal to such mathematical prediction, in agreement with $\mathcal{C}=0$. These results show that the renormalization approach predicts qualitatively the many-particle effects of the exact semi-classical model for stronger regimes.

In order to demonstrate the efficiency of the analytical method in reaching the thermodynamic limit, we present in Fig.\ref{fig1_supp}(b) a single realization of the averaged dipole dynamics for $N=10^5$ atoms. We have used a common desktop computer, equipped with a 6th Generation Intel® Core™ i7 processor, which is capable to run 8 independent tasks in parallel. By calculating 8 temporal points of the dynamics in parallel, from a total of 72 (see Fig.\ref{fig1_supp}(b)), we have got a running time of about 52 hours. For direct numerical integration of Eq.\eqref{sig_nlin}, typically done with Runge Kutta algorithms, such simulations would last at least 7 times more on the same device. Obviously, such discrepancy increases for workstations and clusters. Additionally to time steps parallelization, our analytical solution also allows for calculating the dynamics of each particle separately, even for long-range interacting systems.

For a different initial state, the eigenspectrum selected by the renormalization approach can be qualitatively different. In Fig.\ref{fig2}(a), for example, the case of atoms prepared with a random phase is presented. The absence of correlations between the positions and phases of the atomic dipoles makes the radiation much closer to the single-atom one. As a consequence, the eigenspectrum is centered around the single-atom decay rate. Accordingly, the decay of the total power radiated shows very limited superradiance, which is even weaker in the case of a large excited population ($p_e=0.5$). It happens because cooperative effects in large atomic clouds are a consequence of phase profiles strongly correlated with atomic positions \cite{scully2007}.
\begin{figure}[!h]	
	\begin{center}
		\includegraphics[scale=0.42]{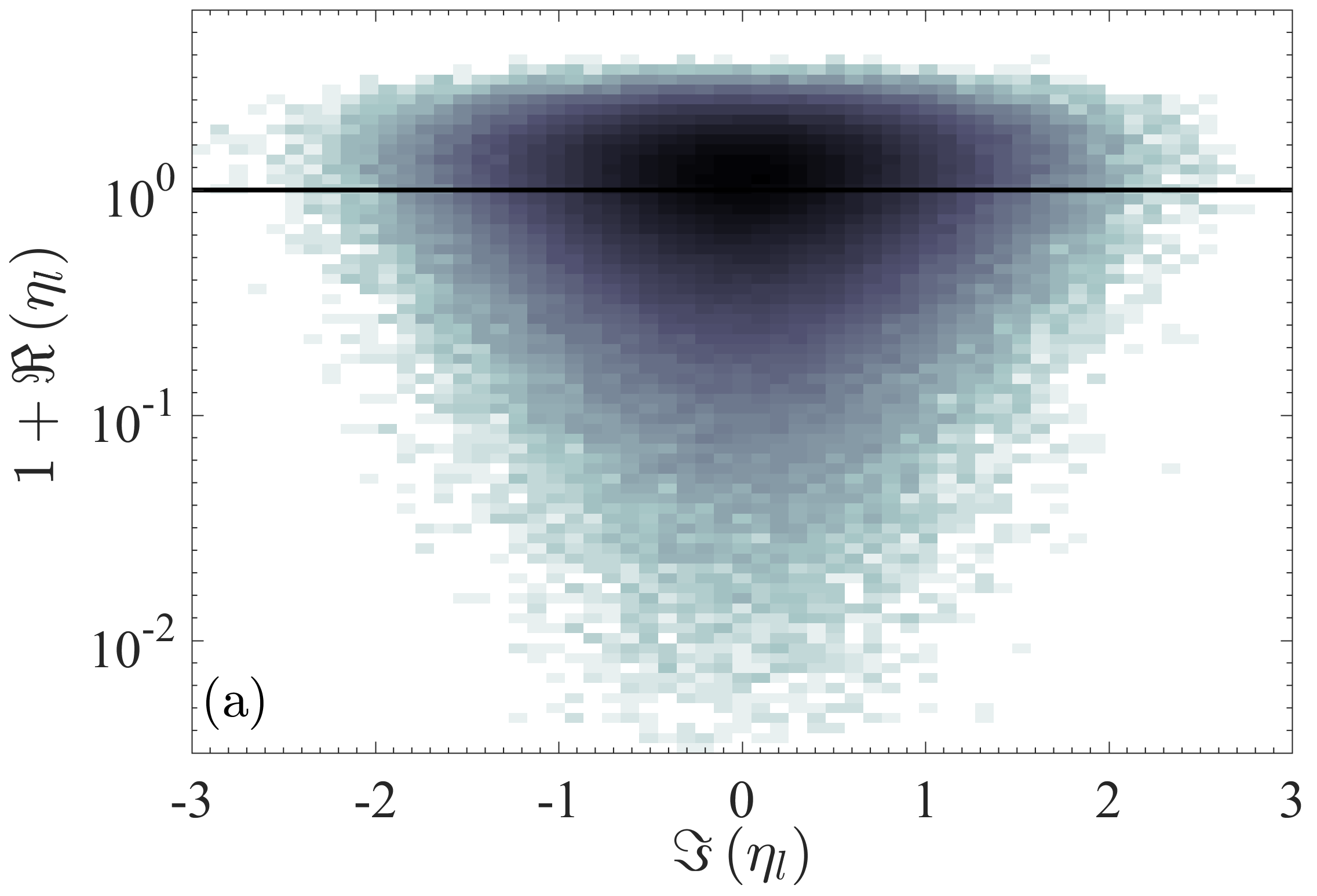}
		\includegraphics[scale=0.41]{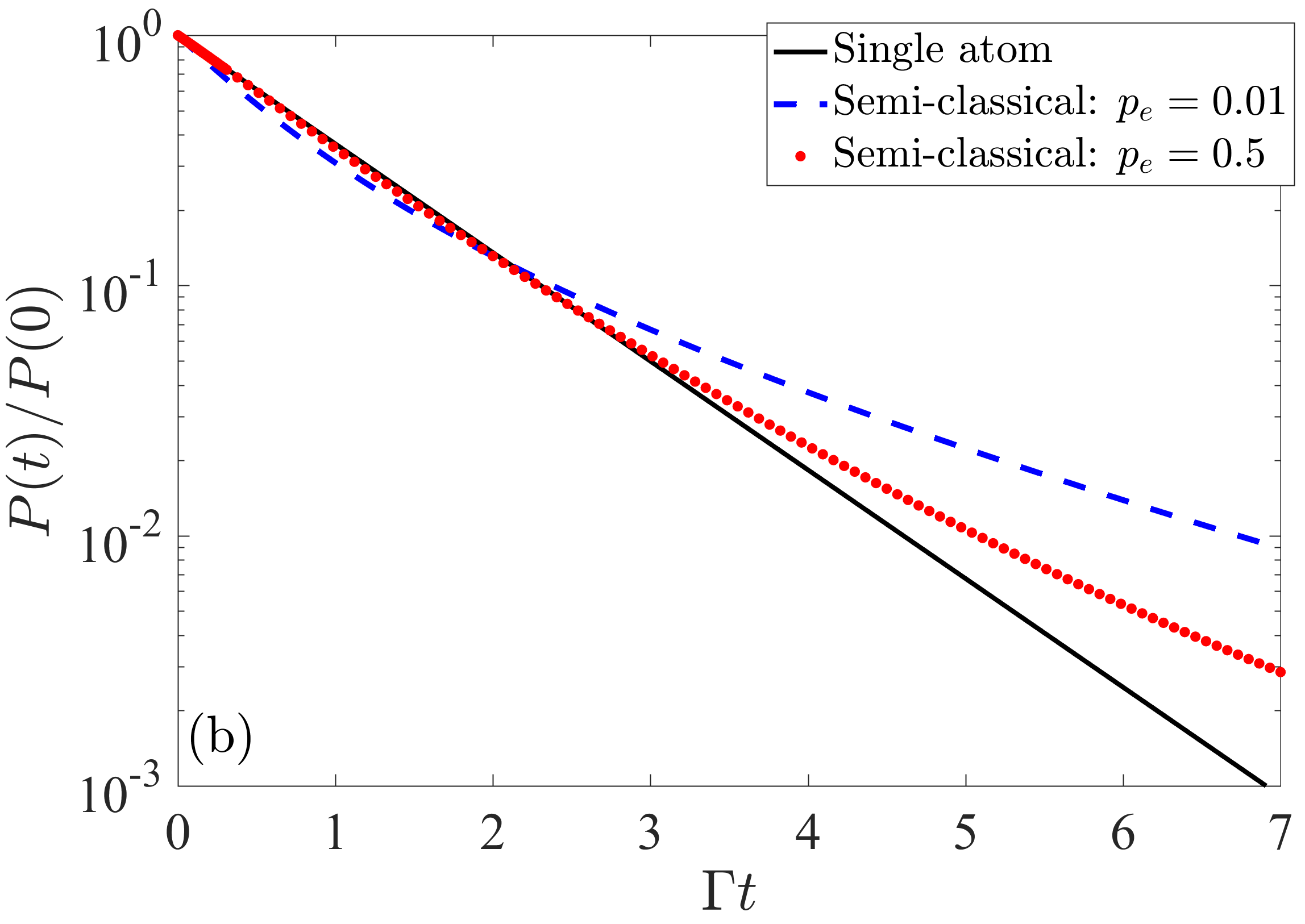}
	\end{center}
	\caption{\label{fig2} (a) Dimensionless effective spectrum predicted by Eq.\eqref{gamma_ren}, for an initial state similar to \eqref{ini}, yet with random phases uncorrelated from the atoms positions. (b) Exact  total radiated power under the same conditions as a function of $\Gamma t$. Simulations realized for $N=25000$ particles and a radius $R=20\lambda$ ($b_0\approx3$); the eigenspectrum corresponds to the case $p_e=0.01$, whereas the radiated power is plotted for $p_e=0.01$ and $0.5$.}
\end{figure}

Another interesting result to point out is the fact that the renormalized decay rates \eqref{gamma_ren} do not depend explicitely on the detuning $\Delta$. This is known feature in linear optics, where the eigenvalues are derived by diagonalization of the coupling matrix underlying the linear dynamics; indeed, the detuning appears there as a diagonal term, so it does not affect directly the eigenspectrum and the associated decay rates. In the renormalization approach, the detuning would appears during the preparation of the system, i.e., in the initial conditions. In this context, a pump largely detuned from the atomic resonance ($(\Delta/\Gamma)^2\gg b_0$) allows one to avoid multiple scattering and reach a state close to \eqref{ini}. Then, despite being in a single-scattering regime, cooperative effects can be observed~\cite{guerin2016,roof2016,kaiser2016}. Our work generalizes this result to the semi-classical dynamics.

\end{document}